\documentclass{JHEP3}

\usepackage{graphicx,amstext,amsmath,amsbsy,amscd,amssymb,epsfig}
\usepackage{latexsym}
\usepackage{cite}

\newcommand{\gthree}{{\gamma_\ast}}

\newcommand{\medsp}{\\[0.7ex]}

\newcommand{\ve}{\varepsilon}

\newcommand{\diff}[1][]{\mbox{d}#1}

\newcommand{\half}[1]{\ensuremath{\frac{#1}{2}}}
\newcommand{\intd}[1]{\int \!\! #1 \;}
\newcommand{\inv}[1]{\ensuremath{\frac{1}{#1}}}

\newcommand{\Stext}[1]{\itindex{\mathcal{S}}{#1}}

\newcommand{\itindex}[2]{\ensuremath{#1_{\mbox{\scriptsize{\itshape #2}}}}}

\DeclareMathOperator{\extdm}{d}
\newcommand{\extd}{\extdm \!}

\newcommand{\B}{Y}
\newcommand{\be}{\begin{equation}}
\newcommand{\ee}{\end{equation}}
\newcommand{\beq}{\begin{equation}}
\newcommand{\eeq}{\end{equation}}
\newcommand{\ba}{\begin{eqnarray}}
\newcommand{\ea}{\end{eqnarray}}

\def\lbldef#1#2{\expandafter\gdef\csname #1\endcsname {#2}}
\def\eqn#1#2{\lbldef{#1}{(\ref{#1})}%
\begin{equation} #2 \label{#1} \end{equation}}

\def\href#1#2{#2}

\newcommand{\ber}{\begin{eqnarray}}
\newcommand{\eer}{\end{eqnarray}}

\newcommand{\beqar}{\begin{eqnarray}}

\newcommand{\eeqar}{\end{eqnarray}}

\newcommand{\eps}{\epsilon}

\newcommand{\dsl}
  {\kern.06em\hbox{\raise.15ex\hbox{$/$}\kern-.56em\hbox{$\partial$}}}

\newcommand{\eeqarr}{\end{eqnarray}}
\newcommand{\ZZ}{{\rm \kern 0.275em Z \kern -0.92em Z}\;}

\newcommand{\Afour}{A} 
\newcommand{\A}{a} 
\newcommand{\killing}{A} 
\newcommand{\Aapp}{A} 

\title{Chemistry of Chern-Simons Supergravity:\\ reduction to a BPS kink, oxidation to M-theory and thermodynamical
aspects} 

\author{Luzi Bergamin\thanks{Supported by project P-16030-N08 of the Austrian Science Foundation (FWF).} , Daniel Grumiller\thanks{Supported by an Erwin-Schr\"odinger fellowship, project J-2330-N08 of the  Austrian Science Foundation (FWF).} , Alfredo Iorio\thanks{Supported in part
by the Physics Department University of Salerno (Italy) and
in part by funds provided by the U.S. Department of Energy (D.O.E.) under cooperative research agreement DF-FC02-94ER40818.}\, and Carlos Nu\~nez\thanks{Supported by a Pappalardo Fellowship and in part by funds provided by the U.S. Department of Energy (D.O.E.) under cooperative research agreement DF-FC02-94ER40818.}\\
\footnotemark[1]\,\,\parbox[t]{13cm}{Institut f\"ur Theoretische Physik, Technische Universit\"at Wien\\ Wiedner Hauptstr.~8-10/136, A-1040 Vienna, Austria} \\ \ \\
\footnotemark[2]\,\,\parbox[t]{13cm}{Institut f\"ur Theoretische Physik, Universit\"at Leipzig\\ Augustusplatz 10-11, D-04103 Leipzig, Germany} \\ \ \\
\footnotemark[3]\,\,\footnotemark[4]\,\,\parbox[t]{13cm}{Center for Theoretical Physics, Massachusetts Institute of Technology\\77, Massachusetts Avenue - Cambridge MA 02139 USA}\\ \ \\
\footnotemark[3]\,\,\parbox[t]{13cm}{Istituto Nazionale di Fisica Nucleare, Rome, Italy}\\ \ \\
E-mail: \email{bergamin@tph.tuwien.ac.at}, \email{grumiller@itp.uni-leipzig.de}, \email{iorio@lns.mit.edu}, \email{nunez@lns.mit.edu}.
}

\abstract{We construct a supersymmetric extension of the two
dimensional Kaluza-Klein-reduced gravitational Chern-Simons term,
and globally study its solutions, labelled by mass and $U(1)$
charge $c$. The kink solution is BPS, and in an appropriate
conformal frame all solutions asymptotically approach $AdS$.
The thermodynamics of the Hawking effect yields interesting
behavior for the specific heat and hints at a Hawking-Page-like
transition at $T_{\rm critical} \sim c^{3/2}$. We address implications
for higher dimensions (``oxidation''), in particular D=3,4 and 11,
and comment briefly on $AdS$/CFT aspects of the kink.}


\preprint{hep-th/0409273\\ TUW--04--27\\ LU--ITP--04/022\\
MIT--CTP--3546}

\dedicated{Dedicated to the people of Iraq}

\begin{document}

\section{Introduction}\label{se:0}

The 3-dimensional (3D) gravity theory governed by the sole
Chern-Simons (CS) term
\begin{equation}\label{cs}
S_{\rm CS} = \frac{1}{4 \pi^2} \int \extd^3 x \epsilon^{\mu \nu
\lambda} \left( \frac{1}{2} \Gamma^\rho_{\mu \sigma}
\partial_\nu \Gamma^\sigma_{\lambda \rho} + \frac{1}{3}
\Gamma^\rho_{\mu \sigma} \Gamma^\sigma_{\nu \tau}
\Gamma^\tau_{\lambda \rho} \right) \;,
\end{equation}
where $\Gamma^\lambda_{\mu \nu} = \frac{1}{2} G^{\lambda \rho}
(\partial_\mu G_{\nu \rho} + \partial_\nu G_{\mu \rho} -
\partial_\rho G_{\mu \nu})$, and $G_{\mu \nu}$ is the 3D
metric tensor\footnote{We will use: $A, B, C,... = 0, 1, 2$ for 3D
flat indices, $\mu , \nu, \lambda, ... = \hat{0}, \hat{1},
\hat{2}$ for the curved ones; while $a, b, c, ... = 0 , 1$ are the
2D flat indices, and $m, n, l, ... = \hat{0}, \hat{1}$ are the
curved ones. When not misleading we will also use upper case
letters for the 3D quantities, like the metric $G_{\mu \nu}$, its
determinant $G$, and lower case for the 2D ones, like the metric
$g_{m n}$, its determinant $g$. But we do not always use this
convention, as for instance, the Dreibein is $e^A_\mu$ and its
inverse $E^\mu_A$. More notation and conventions for spinors are
explained in Appendix A.}, possesses, among others, an interesting
kink solution \cite{Guralnik:2003we}, whose global properties were
extensively investigated \cite{Grumiller:2003ad}. One of the main
goals of this paper is to explore under which circumstances such solutions
are BPS states within a supersymmetric extension and
to connect them with other well studied systems.

The role of the CS term in 3D Einstein-Hilbert (EH) gravity
\begin{equation}\label{eh}
S_{\rm EH} = \kappa \int \extd^3 x \sqrt{G} R \;,
\end{equation}
where $R$ is the 3D scalar curvature, was first studied by Deser,
Jackiw and Templeton (DJT) \cite{Deser:1982wh}. The
DJT action
\begin{equation}\label{djt}
S_{\rm DJT} = S_{\rm EH} + \frac{1}{\mu} S_{\rm CS}
\end{equation}
admits a local excitation $\varphi$ of mass $\mu$, while none
of the two terms separately, $S_{\rm EH}$ or $S_{\rm CS}$,
supports any local excitation due to their entirely topological
nature\footnote{While the topological nature of the CS term is
self-evident, it is more subtle to spot the topological nature of
the EH action in D=3. On this see \cite{Witten:1988hc,Deser:1983tn,vanNieuwenhuizen:1985cx}.}. Thus one might
consider the theory \eqref{cs} as the limiting case ($\mu \to 0$)
of the theory \eqref{djt}.

Solutions of \eqref{cs} were found in \cite{Guralnik:2003we} by
reducing the 3D theory down to two dimensions by means of a
Kaluza-Klein {\it Ansatz}\footnote{In the present case the scalar
field $\varphi$ appears slightly differently from
\cite{Guralnik:2003we}, the difference being a conformal
transformation of the {\em reduced} metric. We will address this
point again below.}
\begin{equation} \label{KK1}
  G_{\mu \nu} = \left( \begin{array}{ccc}
    g_{m n} - \varphi \A_m \A_n &  & -  \varphi \A_m \\
    -  \varphi \A_n &  & - \varphi \
  \end{array}\right) \;,
\end{equation}
where $g_{m n}$ is the D=2 metric tensor, $\A_m$ is a D=2 gauge
vector, and $\varphi$ is a scalar (essentially the conformal
factor). It is assumed that the system has an isometry such that
in the adapted coordinate system implied by \eqref{KK1} all
quantities are independent of one of the coordinates, which will
be denoted by $r$. By expressing the Christoffel connection
$\Gamma^\lambda_{\; \; \mu \nu}$ in terms of the spin connection
$\omega_\mu^{\; A B}$,  the latter---as torsion vanishes---in terms
of Vielbein $e^A_\mu$, using the 3D property $\omega_{\mu \;, A B}
= \epsilon_{A B C} \,\omega_\mu^{\;\; C}$, and $\omega_\mu^{\; A}
= e^B_\mu \omega_B^{\; A}$ we have
\begin{equation}
\omega_A^{\; D}  = \frac{1}{2} \eps^{B C D} \left( e_{\mu A}
\partial_B E_C^\mu + e_{\mu B} \partial_A E_C^\mu + e_{\mu C}
\partial_B E_A^\mu \right) \;,
\end{equation}
where $\eta_{A B} e^A_\mu e^B_\nu = G_{\mu \nu}$, and $e^A_\mu
E^\mu_B = \delta^A_B$. With the {\it Ansatz} \eqref{KK1} the
Dreibein and spin connection read
\begin{equation}\label{dreibein}
e^A_\mu = \left(%
\begin{array}{cc}
  e^a_m & 0 \\
  \sqrt{\varphi} \A_m  & \sqrt{\varphi} \\
\end{array}%
\right) \;,
\end{equation}
\begin{eqnarray}
\omega^a_m & = & \frac{1}{2} e^a_m \sqrt{\varphi} f
+ \frac{1}{2} \A_m \frac{1}{\sqrt\varphi} \epsilon^{a b} E_b^n \partial_n \varphi \;, \\
\omega^2_m & = & - \omega_m - \frac{1}{2} \varphi f \A_m \;, \\
\omega^a_{\hat{2}} & = &  \frac{1}{2} \frac{1}{\sqrt\varphi}
\epsilon^{a b} E_b^m \partial_m \varphi \;, \\
\omega^2_{\hat{2}} & = & - \frac{1}{2} \varphi f \;,
\end{eqnarray}
respectively, where the 2D spin connection $\omega_{m , \; a b} =
\epsilon_{a b} \; \omega_m$ and the dual field strength $f$
\begin{equation}
  \label{eq:deff}
    f_{m n} = \partial_{m} \A_{n} -\partial_n \A_m = \sqrt{- g} \; \epsilon_{m n} f
\end{equation}
have been introduced ($g =
{\rm det}\, g_{m n}$,  recall that $G = {\rm det}\, G_{\mu \nu} =
- \varphi (g)$).

The 2D theory obtained is \cite{Guralnik:2003we}
\begin{equation} \label{csd=2phi}
{\rm CS}  = - \frac{1}{8 \pi^2} \int \extd^2 x \sqrt{-g} \varphi f (r
- \Box \ln \varphi + \varphi f^2) \;,
\end{equation}
where $r = 2 \epsilon^{m n} \partial_m \omega_n$ is the 2D scalar
curvature.

We digress here on the degrees of freedom of this theory.
Off-shell, in D dimensions, a metric $G_{\mu \nu}$ has $D(D+1)/2 -
D = D(D-1)/2$ degrees of freedom, being a symmetric rank-two
tensor and the action being invariant under D reparametrizations.
This gives 3 degrees of freedom in D=3 (and 1 in D=2). The theory
\eqref{cs} also enjoys conformal invariance\footnote{That is why,
sometimes, in literature the Chern-Simons gravitational term is
also referred to as ``conformal gravity'' \cite{Horne:1988jf}.} as
a consequence of the eqs.~of motion which imply the vanishing
of the Cotton tensor \eqn{cottontensor}{2\sqrt{g}C^{\mu\nu}=
\epsilon^{\mu\sigma\tau} D_\sigma R^{\nu}_\tau +
\epsilon^{\nu\sigma\tau} D_\sigma R^{\mu}_\tau\,, } where $D_\mu$
is the covariant derivative based on $\omega^{AB}_\mu$. For this
3D action the true degrees of freedom off-shell are only 2 as we
can get rid of the conformal factor $\varphi$ by setting it to 1.
This can be seen by performing a conformal transformation of the
{\em reduced} theory, $g_{mn}\to\varphi g_{mn}$, assuming positive
$\varphi$ (as the quantity $\varphi$ has to be nonvanishing for a
non-degenerate metric \eqref{KK1} this is not an additional
restriction). As the gauge field $\A_m$ remains unchanged, the
field strength $f$ behaves inversely to the volume form ~$f\to
f/\varphi$. With the well-known relations
\begin{equation}
  \label{eq:ct}
  \sqrt{-g}\,r\to\sqrt{-g}\,r+\sqrt{-g}\,\square\ln{\varphi}
\end{equation}
and $\sqrt{-g}\to\varphi\sqrt{-g}$ one obtains from \eqref{csd=2phi} the simpler action
\begin{equation} \label{csd=2}
{\rm CS}  = - \frac{1}{8 \pi^2} \int \extd^2 x \sqrt{-g}  \; (f r +
f^3) \;,
\end{equation}
leaving 1 degree of freedom for $r$ and 1 for $\A_m$ (recall that
the performed reduction does not change the total number of
degrees of freedom, but simply rearranges them). Obviously, the
scalar field $\varphi$ can play no role as the action
\eqref{csd=2} is {\em independent} thereof.\footnote{We thank
Roman Jackiw for helpful discussions on this point.} On the other
hand, conformal invariance does not hold in the EH
sector\footnote{Nonetheless, it is customary in standard
Kaluza-Klein reduction of the EH term to set $\varphi =1$ {\it in
the action}. This is to avoid a meaningless theory by setting
$\varphi =1$ in the Euler-Lagrange eqs. On this point see,
for instance, \cite{O'Raifeartaigh:2000vm}.} as can be seen easily by reducing the
action (\ref{eh}) with the \emph{Ansatz} (\ref{KK1})
\begin{equation} \label{ehd=2}
{\rm EH} \sim \int \extd^2 x \sqrt{-g} \sqrt{\varphi} (r + \varphi \frac{1}{2}
f^2) \;,
\end{equation}
hence the degrees of freedom off-shell are three: one for $r$,
one for $\A_m$, and one for $\varphi$.

As well known, this discussion is valid {\it off-shell}, as {\it
on-shell} neither CS in \eqref{cs} nor EH in \eqref{eh},
separately, have any degree of freedom. It is interesting to note
that classically a conformal transformation on the 2D metric is
always possible and this gives another perspective on the origin
of the dynamical (on-shell) degree of freedom of DJT: by virtue of
\eqref{eq:ct} the kinetic term for $\ln \varphi$ can never
disappear when {\it both} terms \eqref{csd=2phi} and \eqref{ehd=2}
are considered. We might call this mode a ``conformal mode''.

In \cite{Guralnik:2003we} solutions of \eqref{csd=2} were found,
with field strengths
\begin{equation}
f = 0 \quad {\rm (A)} \;, \quad f = \pm \sqrt c \quad {\rm (B)}
\;, \quad f = \sqrt c \tanh \frac{\sqrt c}{2} x \quad {\rm (C)}
\label{fsol} \;,
\end{equation}
corresponding scalar curvatures
\begin{equation}
r = c \lessgtr 0 \quad {\rm (A)} \;, \quad r = - 2 c < 0 \quad
{\rm (B)} \;, \quad r = - 2 c + 3 c/ \cosh^{2} \frac{\sqrt c}{2} x
\quad {\rm (C)} \label{rsol} \;,
\end{equation}
and line elements
\begin{eqnarray}
(\extd s)^2 & = & \frac{2}{c t^2} [(\extd t)^2 - (\extd x)^2] \quad c > 0
\quad {\rm (A)} \;, \label{ds1} \\
(\extd s)^2 & = & \frac{2}{|c| x^2} [(\extd t)^2 - (\extd x)^2] \quad c < 0
\quad {\rm (A)} \;, \label{ds2} \\
(\extd s)^2 & = & \frac{1}{c x^2} [(\extd t)^2 - (\extd x)^2] \quad c > 0
    \quad {\rm (B)} \;, \label{ds3} \\
(\extd s)^2 & = & \inv{\cosh^{4} \frac{\sqrt c}{2} x} (\extd t)^2 - (\extd x)^2 \quad
c > 0   \quad {\rm (C)} \;. \label{ds4}
\end{eqnarray}
By using a suitable coordinate system the (kinky) spacetime in (C)
may be extended beyond $x=\pm\infty$, as the affine distance to
this ``boundary'' is finite (see \cite{Grumiller:2003ad} and
~Eq.~\eqref{eq:cs6} below).

In the following Sections we intend to construct a
suitable supersymmetric theory compatible with (at least some)
supersymmetry (SUSY) of the line elements \eqref{ds3} and \eqref{ds4}. The paper
is organized as follows: Section 2 is devoted to the construction,
via  the rigorous methods of graded Poisson Sigma Models (gPSM),
of the suitable SUSY for the 2D model \eqref{csd=2}. There we
present our supersymmetric CS (SUCS) 2D Lagrangian and
transformations, and we show the differences with the theory
obtained by reducing the 3D SUCS. In Section 3 we study the SUSY
of the solutions and discuss their global properties. In Section 4
we compare our 2D model to the 3D model of \cite{Deser:1983sw}
and we spell a connection with 4D supergravity and M theory. Finally Section 5 contains our
conclusions. Appendix A explains notation and conventions of the 2D formulation,
Appendix B provides details on the formulation of SUCS as a gPSM,
and Appendix \ref{app:C} discusses conformal transformations of
the reduced theory.

\section{Construction of the 2D SUGRA model}\label{se:1}

To identify the conditions under which the metrics \eqref{ds3} and
\eqref{ds4} preserve some SUSY we will follow the usual
prescription of finding covariantly constant local SUSY parameters
$\epsilon_\alpha (x)$ (see for example \cite{Edelstein:1993bb,Edelstein:1996md}) satisfying
\begin{equation}\label{Deps=0}
\hat{D}_\mu \epsilon_\alpha = (D_\mu+ F_\mu) \epsilon_\alpha =
\partial_\mu \epsilon_\alpha + \omega_\mu^{A B} \left( \Gamma_{A
B} \right)_\alpha^\beta \epsilon_\beta + (F_\mu\epsilon)_\alpha =
0 \;,
\end{equation}
where $D_\mu$ is the covariant derivative based on
$\omega_\mu^{A B}$, $\Gamma_{A B}$ are the local Lorentz
generators,  $F_\mu$ is a one form that will be determined below
and is related to additional fields present in the system. We have
not specified yet the dimension D of the spacetime. The general
idea is that on a curved manifold it is not possible in principle
to define such spinors, because when parallel transported, the
spinor transforms as \beq \epsilon(x)= e^{i\int_0^x
\omega}\epsilon(0) \;, \label{paralell} \eeq so, in a closed path
$\epsilon(0)= e^{i\oint \omega}\epsilon(0)$, in general the spinor
is not uniquely defined. But it might happen that other terms do
contribute to the parallel transport, such that $\epsilon(0)=
e^{i\oint (\omega + F)}\epsilon(0)$, being $F$ the one form
related to $F_\mu$ above.
When a cancellation between $\omega$ and $F$ occurs, the spinors
are compatible with (some fraction of) SUSY, hence the given field
configuration is (partially) supersymmetric.

We will be interested in purely bosonic configurations with vanishing
fermionic variations, as bosonic SUSY variations will vanish trivially
in these cases (generalizations of this concept are discussed briefly below). This is why it is usually
stated that a bosonic field configuration is supersymmetric if the
fermionic variations vanish. Examples of this are all the extremal
$Dp$ branes in type II SUGRA theories and extremal M-branes in
D=11 SUGRA.

Another way of looking at Eq. \eqref{Deps=0} is to consider the
supergravity (SUGRA) transformations
\begin{align}
  \delta {e_\mu}^A &= - 2i (\ve \Gamma^A \psi_\mu)\ , \label{sugra1}\medsp
  \label{sugra2}
  \delta \psi_{\mu \alpha} &= - \hat{D}_\mu \ve_{\alpha} \;,
\end{align}
in the case when the Rarita-Schwinger fermions $\psi_{\mu
\alpha}$ (partners of the bosonic Vielbein ${e_\mu}^A$) and their
variations vanish \cite{Edelstein:1996md}. In this case Eq.
\eqref{Deps=0} is a consequence of Eq. \eqref{sugra2}. For the 3D
DJT model a SUGRA extension is available \cite{Deser:1983sw}, and the SUGRA multiplet is as given in
\eqref{sugra1}-\eqref{sugra2}.

\subsection{The SUGRA multiplet}

We will now exploit these methods and propose the following SUGRA
multiplet and transformations for the 2D theory \eqref{csd=2}
\begin{align}
  \delta {e_m}^a &= - 2i (\ve\gamma^a \psi_m)\ , \label{eq:scs12.3main}\medsp
  \label{eq:scs12.6main}
  \delta \A_m &= - 2\ve \gamma_* \psi_m\ ,\medsp
  \label{eq:scs12.4main}
  \delta \psi_{m \alpha} &= - \hat{D}_m \ve_\alpha
\end{align}
where
\begin{equation}\label{hatD}
    \hat{D}_m \ve_{\alpha} = \partial_m \ve_\alpha + \frac{1}{2} \tilde{\omega}_m (\gamma_* \ve)_\alpha +
    \half{i} F (\gamma_m \ve)_\alpha \;,
\end{equation}
\begin{equation} \label{superf}
F = f + \epsilon^{mn} \psi_n \gamma_* \psi_m \;.
\end{equation}
In D=2 $\frac{1}{2} \gamma_*$ is the generator of Lorentz
transformations in spinor space, $\gamma_m = e^a_m \gamma_a$, and
the torsionfull connection reads ($\tilde{\omega}_m = e^a_m
\tilde{\omega}_a$, more details on the notation are explained in Appendix \ref{app:A})
\begin{equation}\label{torspinconn}
    \tilde{\omega}_a = \epsilon^{mn} \partial_n e_{ma} - i
\epsilon^{mn} (\psi_n \gamma_a \psi_m) \;.
\end{equation}

Immediately one sees that the balance of fermionic and bosonic
degrees of freedom is correct: the 2 bosonic degrees of freedom (1
for $e^a_m$ and 1 for $\A_m$) are transformed into 2 fermionic
degrees of freedom ($\psi_{m \alpha}$ has 4 independent components
to which we must subtract 2 components as consequence of the
irreducibility condition $\gamma_m \psi^m = 0$, leaving 2
fermionic degrees of freedom).

The transformation of the Zweibein is customary, as well as the
transformation of the Rarita-Schwinger field (when the previous
argument on the ``improved'' covariant derivative is assumed),
although the exact expression for the extra term $\half{i} F
(\gamma_m \ve)_\alpha $ will be fully justified below in the gPSM
construction. What is more subtle is the transformation law for
the gauge field, that might look rather unusual. Our qualitative
argument for it is based on the conformal symmetry of the
Chern-Simons term (\ref{cs}): from the Kaluza-Klein split of the
Dreibein \eqref{dreibein} and the SUGRA transformations
\eqref{sugra1} one sees that the components $e^2_m =
\sqrt{\varphi} \A_m$ transform into something proportional to $\ve
\gamma_* \psi_m$. This would be true also for the EH sector of the
DJT theory, but for the CS sector we can invoke conformal symmetry
and set $\varphi =1$.

To prove that our {\em Ansatz}
\eqref{eq:scs12.3main}-\eqref{superf} is viable we must now
supersymmetrize \eqref{csd=2}. This could be done in different
ways, for instance using the metric as the basic geometric
variable (second order formalism). We will, instead, employ as
basic objects the Zweibein and connection (first order formalism),
as for dilaton (super-)gravity in D=2 this approach provides
powerful tools to discuss classical and quantum aspects of the
theory (for a review cf.~\cite{Grumiller:2002nm} and
refs.~therein). Once this formalism is incorporated classical
solutions may be obtained {\em globally} with particular ease.
These successes are intimately related to the fact that it is
essentially a special case of a
(graded) Poisson Sigma Model 
(gPSM)
\cite{Schaller:1994es},
that we will now briefly introduce (for a more comprehensive
review cf.~e.g.~\cite{Schaller:1994uj}).

\subsection{gPSM supersymmetrization: the action}

The action of a gPSM reads\footnote{Geometrically, there is a 2D
base manifold $\mathcal{M}_2$ and a target space $\mathcal{N}$,
the latter being a Poisson manifold with associated Poisson tensor
$P^{IJ}$ and with coordinates denoted by $X^I$ (indices
$I,J,K,\dots$ run from 1 to the dimension of the target space).
Those coordinates as well as the gauge fields $A_I$ are functions
of the coordinates $x^m$ on the base manifold, $X^I(x)$ and
$A_I(x)$.  The same symbols are used to denote the mapping of
$\mathcal{M}_2$ to $\mathcal{N}$. The $\extd X^I$ stand for the
pullback of the target space differential $\extd X^I=\extd
x^m\partial_m X^I$ and $A_I$ are 1-forms on $\mathcal{M}_2$ with
values in the cotangent space of $\mathcal{N}$. It is sometimes
convenient to interpret the gauge field as a 1-form not only with
respect to the base manifold but also with respect to the target
space, $\mathcal{A}=\extd X^I \wedge A_I$. It has been shown that
a gPSM arises as the most general consistent deformation ({\em \`a
la} Barnich and Henneaux \cite{Barnich:1993vg}) of 2D BF theory
\cite{Izawa:1999ib}.}
\begin{equation}
  \label{eq:gPSM}
  {\mathcal S}_{gPSM}=\int_{\mathcal{M}_2} \extd X^I\wedge A_I + \frac12 P^{IJ} A_J\wedge A_I\,.
\end{equation}
The name ``Poisson Sigma Model'' is justified because it is a
Sigma model (the target space coordinates behave as scalar
fields from the world-sheet point of view) and because the target
space is a Poisson manifold. ``Graded'' refers to the fact that
one allows for a $\mathbb{Z}_2$ grading of the target space in order to accommodate
SUSY. Thus, a gPSM is defined by the specification of the Poisson
tensor $P^{IJ}$ depending on the target space coordinates $X^K$:
its dimension, its maximal rank, its $\mathbb{Z}_2$ grading
properties and its entries. For theories describing gravity
generically it has a nontrivial kernel and thus it is not
symplectic.

As a consequence of the graded non-linear Jacobi identities
\begin{equation}
\label{eq:nijenhuis}
  P^{IL}\partial _{L}P^{JK}+ \mbox{\it
g-perm}\left( IJK\right) = 0\,,
\end{equation}
where $\mbox{\it g-perm}$ stands for all graded cyclic
permutations, a gPSM \eqref{eq:gPSM} is invariant under the
nonlinear symmetry transformations with symmetry parameters
$\ve_I(X^J, x^m)$
\begin{align}
\label{eq:symtrans}
  \delta X^{I} &= P^{IJ} \ve_{J}\ , \medsp
  \delta A_{I} &= -\mbox{d} \ve_{I}-\left( \partial _{I}P^{JK}\right) \ve _{K}\, A_{J}\,,\label{eq:symtrans2}
\end{align}
where $\partial_I:=\partial/\partial X^I$. The symmetries
(\ref{eq:symtrans}), (\ref{eq:symtrans2}) comprise local Lorentz
transformations, diffeomorphisms and local SUSYs, respectively.

The identities \eqref{eq:nijenhuis} pose non-trivial constraints
on the possible form of the Poisson tensor $P^{IJ}$. If $P^{IJ}$
is linear the otherwise non-linear gauge transformations
\eqref{eq:symtrans2} reduce to ordinary non-abelian ones; the term
$\partial_I P^{JK}$ yields the structure constants which also
enter the eqs.~of motion. They are obtained by varying
\eqref{eq:gPSM} with respect to $X^I$ and $A_I$, and are of first
order. A simple (Hamiltonian) analysis shows that the number of
physical propagating field degrees of freedom is zero. In this
sense, the theory is a topological field theory of Schwarz type,
cf.~\cite{Birmingham:1991ty} for a review.

The absence of propagating physical modes does not necessarily
imply that the model is trivial. In the case of gravity the
non-trivial features are encoded in the number and
types of (Killing) horizons and singularities as well as in the
asymptotic behavior and the properties of geodesics of test
particles (locally the model indeed is trivial as every 2D metric
is conformally flat in a certain patch).

To derive the gPSM of interest in the present case we can rely on the
knowledge of such models describing $N=(1,1)$
dilaton SUGRA.\footnote{It should be emphasized, as discussed in
refs.~\cite{Bergamin:2002ju,Bergamin:2003am}, that not every gPSM
which in its bosonic limit reduces to dilaton gravity
\cite{Ertl:2000si} can be considered as genuine dilaton SUGRA
\cite{Park:1993sd}. Thus, dilaton SUGRA is not merely a gPSM, but
a gPSM with a little bit of extra structure -- this should not
come as a surprise because also in the bosonic theory one needs
extra structure for a PSM to be a (first order) dilaton gravity
model.} They require three bosonic (the ``dilaton'' $X$ which
can be interpreted as Lagrange multiplier for curvature and the
Lagrange multipliers for torsion $X^a$) and two fermionic (the
``dilatini'' $\chi^\alpha$) target space coordinates. It is
emphasized that this target space is quite different from the
``standard target space'' of string theory. In particular, as
mentioned before, one of the target space coordinates corresponds
to what in the string literature is referred to as ``dilaton
field'' (cf.~footnote \ref{fn:st}).

There are only two free functions of the dilaton field $X$
defining the model.\footnote{For the reader familiar with scalar
tensor theories it may be helpful to note that these two free
functions of the target space coordinate $X$ are {\em equivalent}
to the ones appearing in the bosonic second order action
\[\int_{\mathcal{M}_2} \extd^2x \sqrt{-g}\left(XR-Z(X)(\nabla X)^2+2V(X)\right)\,.\]
Note also that in non-gPSM literature often an exponential
representation of the dilaton field is used, $X=e^{-2\Phi}$.
\label{fn:st}} Appealing to conformal transformations eliminates
one of these and thus, neglecting subtleties with singular
conformal transformations and quantum inequivalence, one has to
specify only the prepotential $u(X)$ to be defined below. Thus,
the art is to find the correct prepotential. Once this is achieved
standard methods can be applied to solve the model locally and
globally.

To each target space coordinate $X^I$ corresponds a gauge field
$A_I$: The bosonic spin-connection $\omega$ and Zweibeine $e_a$;
the fermionic gravitini $\psi_\alpha$. To describe the bosonic
sector of SUCS \cite{Guralnik:2003we} we need an extra bosonic
target space coordinate \cite{Grumiller:2003ad} which we denote
here by $\B$. The corresponding abelian gauge field 1-form is
denoted by $\Aapp$. The prepotential may thus depend not only on
$X$ but also on $\B$.

It is emphasized again that once the prepotential is chosen
everything is fixed and all classical solutions may be obtained
globally. Without further ado we present the correct prepotential
\begin{equation}
  \label{eq:scs1}
  u = X^2 - \B \;,
\end{equation}
and explain in Appendix \ref{app:B} why it is the correct one and
how it enters the Poisson tensor. Exploiting the relation between
prepotential and potential \cite{Ertl:2000si},
\begin{equation}
  \label{eq:scs1a}
  V=-\frac18 \frac{\diff}{\diff X} u^2 = \frac12 (X \B -X^3)\,,
\end{equation}
Eq.~(\ref{eq:scs1a}) is seen to be equivalent to Eq.~(3) in
\cite{Grumiller:2003ad} without any obstruction.\footnote{This is
not always the case as obviously not every potential $V(X,\B)$ may
be expressed in terms of a prepotential $u(X,\B)$.} This is
necessary for a meaningful description of SUCS but not sufficient
yet; the missing information is contained in the gauge
transformations discussed in Appendix \ref{app:B} which is
recommended for the reader interested in the relevant steps
performed in the first order formulation. As crucial result $A$
does not acquire a SUSY partner and does not transform as a gauge
field (cf.\ eq.\  \eqref{eq:scs12.6}), but rather as a component
of the Dreibein, as expected from the theory in D=3. This is quite
non-trivial as we have adjusted the prepotential solely to
establish the correct bosonic potential, but nonetheless the
``miracle'' happens and it provides also the transformation law
for $\Aapp$ we proposed in \eqref{eq:scs12.6main}.

The main results of Appendix \ref{app:B} are the first order
formulation of SUCS -- which then allows to apply the discussion
of \cite{Bergamin:2003mh} to study BPS solutions -- and the
transformation to the classically equivalent second order
formulation
\begin{equation}
  \label{eq:scs13main}
{\cal S}^I_{\rm SUCS} = -\frac{1}{8\pi^2} \intd{\diff{^2
x}\sqrt{-g}} \Bigl( \tilde{r} F + F^3 + \Sigma^2 - F^2 \Delta
\Bigr)
\end{equation}
with $\tilde{r} = 2 \epsilon^{mn} \partial_m \tilde{\omega_n}$,
$\Sigma_\alpha = 2 \epsilon^{n m} \hat{D}_m \psi_{n \alpha}$,
$\Delta=\epsilon^{mn}\psi_n\gthree\psi_m$, and with the SUSY
transformations \eqref{eq:scs12.3main}--\eqref{eq:scs12.4main}.

It is interesting to note that the action \eqref{eq:scs13main} can be obtained from
superspace techniques as well\footnote{We thank the referee for drawing our attention to this
point.}. As our further calculations do not rely on
superspace formulations we simply present the final result at this place and
outline the derivation in Appendix \ref{app:B}. Two-dimensional supergravity
in superspace \cite{Howe:1979ia} is formulated in terms of a scalar superfield
$S$ with field content $(e_m^a, \psi_m^\alpha, \mathcal{A})$. $\mathcal{A}$ is
the auxiliary field and appears as lowest component in $S$. The action
\eqref{eq:scs13main} is equivalent to the superspace action ($E$ denotes the
superdeterminant of the super-Vielbein)
\begin{equation}
  \label{eq:howe1}
  {\cal S}_{\rm SP} = \intd{\diff{^2x \extd^2 \theta}} E S^2
\end{equation}
if the auxiliary field is identified with the $\mathcal{A} = -F$.

\subsection{A different supersymmetrization}

Although the following Sections will be devoted to the study of
the model just constructed,  we want to mention now a different (if not more direct)
strategy that is to reduce the 3D SUCS action and transformations
available in literature \cite{Deser:1983sw}. We will present here
the results of such a reduction, show that it is not equivalent to
the model just constructed, and leave its study to a future work.

From the Kaluza-Klein {\em Ansatz} \eqref{KK1} with $\varphi=1$
and the SUSY variation \eqref{sugra1} it follows that one has to
choose $\psi_{\hat{2} \alpha} = 0$ in accordance with the
discussion above. Then the reduced supersymmetric spin-connection
is equivalent to the bosonic result (1.7)-(1.10) if the 2D
spin-connection $\omega_m$ and $f$ are replaced by their SUSY
covariant versions $\tilde{\omega}_m$ in \eqref{torspinconn} and
$F$ in \eqref{superf}, respectively. The SUSY transformations of
the reduced bosonic variables $e_m{}^a$ and $a_m$ coincide with
\eqref{eq:scs12.3main} and \eqref{eq:scs12.6main}, while the
transformations of the spinors become
\begin{align}
\label{eq:luzi1}
    \delta \psi_{m\alpha} &= - \hat{D}_m \ve_\alpha + \frac{1}{4} F \bigl(
    i (\gamma_m\ve)_\alpha - a_m (\gthree \ve)_\alpha \bigr)\ , \medsp
\label{eq:luzi2}
    \delta \psi_{\hat{2} \alpha} &= \inv{4} F (\gthree \psi_{\hat{2}})_\alpha \ .
\end{align}
In contrast to the supersymmertization discussed above, the conformal frame defined by $e_{\hat{2}}{}^2 = 1$, $\psi_{\hat{2}}=0$ is not supersymmetric, as can be seen from \eqref{eq:luzi2}. Thus a suitable Wess-Zumino gauge, where the SUSY
transformations are combined with an appropriate conformal
transformation, should be implemented.

With this in hand, the reduction of the 3D SUCS action is
straightforward. In the notation of \eqref{eq:scs13main} the
result can be written as
\begin{eqnarray}
  \label{eq:luzi3}
  {\cal S}^{II}_{\rm SUCS} &=& - \inv{8 \pi^2} \intd{\diff{^2 x}} \sqrt{-g} \Bigl(\tilde{r}
  F + F^3 + \Sigma^2 + 2i F (\Sigma \gamma^m \gthree \psi_m) - \half{1} F^2 \psi^m
  \psi_m \Bigr) \\
  &=& - \inv{8 \pi^2} \intd{\diff{^2 x}} \sqrt{-g} \Bigl(\tilde{r}
  F + F^3 + \Sigma^2 - \half{1} F^2 \Delta \Bigr) \label{eq:luzi4}
  \;,
\end{eqnarray}
where \eqref{eq:luzi4} is obtained from \eqref{eq:luzi3} once the
irreducibility constraint for Rarita-Schwinger fields, $\psi^a
\gamma_a=0$, is implemented\footnote{This constraint is not
supersymmetric for the model at hand and a WZ gauge should be
used. However, it is used for illustrational purposes
only and is never implemented in the remainder of this paper.}.

The theory \eqref{eq:luzi4} with transformations
\eqref{eq:scs12.3main}, \eqref{eq:scs12.6main} and
\eqref{eq:luzi1} is not equivalent to the theory
\eqref{eq:scs13main} with transformations
\eqref{eq:scs12.3main}-\eqref{eq:scs12.4main}, as the fermionic
transformations and potential differ.

\section{SUSY of the solutions}

We now have all the ingredients to see under which conditions the
solutions of the 2D theory are supersymmetric.
According to Eq. \eqref{eq:scs12.4main} we are looking for solutions of
\begin{equation}\label{dpsiD=2}
\delta \psi_{m \alpha} = - \partial_m \epsilon_\alpha +
\frac{1}{2} \omega_m \left(\gthree\right)_\alpha^{\; \; \beta}
\epsilon_\beta - i \frac{f}{2} e_m^{\; a}
\left(\gamma_a\right)_\alpha^{\; \; \beta} \epsilon_\beta = 0 \;,
\end{equation}
where only the bosonic contributions $f$ and $\omega_m$ (to $F$
and $\tilde{\omega}_m$, resp.) are important. We find that: no
SUSY exists for the line elements \eqref{ds1} and \eqref{ds2},
while \eqref{ds3} and \eqref{ds4} support half SUSY. We outline
these results in some details, starting from the important case of
the kink \eqref{ds4}.

In a more general setting, where the
line element \eqref{ds4} is written in the form (note our signature $(+,-)$)
\begin{equation}\label{2metric}
(\extd s)^2= \killing(x)^2 (\extd t)^2 - (\extd x)^2 \;.
\end{equation}
The only nonzero component of the spin-connection for this metric
is $\omega_{\hat{0}} = \killing'$, hence the $m= \hat{1}$ component of
Eq. \eqref{dpsiD=2} is
\begin{equation}\label{gravi1}
\partial_x \epsilon +\frac{i}{2} f (x) \gamma_{1} \epsilon
= 0 \;,
\end{equation}
and the $m = \hat{0}$ component is
\begin{equation}\label{gravi2}
\partial_t \epsilon - \frac{\killing'(x)}{2} \gthree \epsilon + \frac{i}{2}\killing(x) f (x) \gamma_{0} \epsilon
= 0 \;,
\end{equation}
where the $\gamma$ matrices are representation independent and
written in flat indices (note that $\gamma_0\gamma_1=\gthree$).
We can assume that the spinor (like
the full configuration) is time independent. We also need to
impose one projection on the spinor, this preserves just half SUSY
\begin{equation}\label{proj}
{\gamma_{1} \epsilon = -i \rho \epsilon, \;\;\; \rho=\pm 1 \;.}
\end{equation}
Then the solutions of \eqref{gravi2} are
\begin{equation}\label{solgrav1}
{\epsilon(x) = e^{-\rho / 2 \int\limits^x_0 \extd x' f(x')} \epsilon(0)}
\end{equation}
with,
\begin{equation}\label{solgrav2}
{ \frac{\rho \killing'}{\killing} = f}
\end{equation}
and $\epsilon(0)$ being a constant spinor satisfying the projection
(\ref{proj}). We see that the kink falls in this class, with $\killing(x)$ and $f(x)$ given in \eqref{ds4} and in \eqref{fsol}-(C),
respectively, because it satisfies the constraint (\ref{solgrav2})
with $\rho = -1$, explicitly
\begin{equation}\label{solgrav3}
    \epsilon (x) = \epsilon(0)/ \cosh \frac{\sqrt c}{2} x \;.
\end{equation}
According to (\ref{paralell}) and the formula in the text below this
can also be written as
\beq \epsilon(x)= e^{i\int\limits_0^x (\omega - F)}\epsilon(0)\,,
\label{paralell2} \eeq
where $F$ is the matter contribution to the parallel transport given by the third term in Eq.~\eqref{dpsiD=2}.

The symmetry breaking solutions $f = \pm \sqrt{c}$ (\eqref{fsol} (B)
and line element \eqref{ds3}) also admit
1/2 SUSY.
The last two cases, the line elements \eqref{ds1} and \eqref{ds2},
show no SUSY because one of the two eqs.~\eqref{dpsiD=2}
requires $\epsilon_\alpha$ {\it independent} from $t$ (from $x$)
for $c > 0$ (for $c < 0$), while the other equation requires that
$\epsilon_\alpha$ has to {\it depend} on $t$ (on $x$), leading to
a contradiction.

As the solutions \eqref{ds1}-\eqref{ds4} were found in
\cite{Guralnik:2003we} by a local analysis and by tuning one of
the two constants of motion to a definite value, the
considerations above are valid only in this sense. However, the
solutions may be extended globally \cite{Grumiller:2003ad}, and we
will show this in the next subsection in particular for those preserving 1/2 SUSY.

\subsection{All classical solutions}

\subsubsection{Bosonic part}

For the convenience of the reader not so familiar with dilaton gravity \cite{Grumiller:2002nm} we recall briefly how to obtain bosonic solutions locally in the first order formalism and how to extend them globally by means of geodesic extension (for more details and further refs.~cf.~\cite{Grumiller:2003ad}). To this end one has to solve the eqs.~of motion \eqref{eq:gPSMeom1}, \eqref{eq:gPSMeom2}, which read explicitly (dropping all wedges)
\begin{gather}
  \extd X - X^b {\epsilon_b}^a e_a + \half{1} (\chi \gthree \psi ) = 0
  \label{eq:gPSMeom3.1}\ , \medsp
  D X^a + \frac12 \epsilon^{ab} e_b \bigl(X\B-X^3  + \frac14 \chi^2
  \bigr) + \frac{i}{2}X(\chi \gamma^a \psi) = -W^a \label{eq:gPSMeom3.2}\ ,\medsp
  D \chi^\alpha - \frac{i}{2}X(\chi \gamma^a)^\alpha e_a + 2 i X^a (\psi
  \gamma_a)^\alpha - (X^2-\B) (\psi\gthree)^\alpha  = 0
  \label{eq:gPSMeom3.3}\ ,\medsp
  \extd Y=0\label{eq:neweom1}\ , \medsp
  \extd \omega + \frac12 \epsilon \bigl( \B-3X^2 \bigr) -  \frac{i}{2}
  (\chi \gamma^a e_a \psi) - X (\psi \gthree \psi) = -W
  \label{eq:gPSMeom4.1}\ ,\medsp
  D e_a + i (\psi \gamma_a \psi) = 0 \label{eq:gPSMeom4.2}\ ,
  \medsp
  D \psi_\alpha +\frac14 \epsilon \chi_\alpha - \frac{i}{2} X (\gamma^a e_a \psi)_\alpha = -W_\alpha
  \label{eq:gPSMeom4.3}\ , \medsp
  \extd\Aapp+\frac12 \epsilon X+\frac12 \psi\gthree\psi = 0\label{eq:neweom2}\ .
\end{gather}
We recall the definition of the 0-form fields: $X,X^a,\chi^\alpha,\B$ are dilaton, Lagrange multipliers for torsion, dilatino and $U(1)$ charge, respectively, and the gauge field 1-forms: $\omega,e_a,\psi_\alpha,A$ are spin connection, dyad (related to the Zweibein), gravitino and $U(1)$ connection, respectively. In the absence of matter the energy-momentum 1-forms $W^a=W_\alpha=0$ vanish identically; the 2-form $W$ is zero as well in that case, but it vanishes already for minimal coupling, i.e., if the matter action does not depend on the dilaton $X$. For a given matter action $S_{(m)}$ in light-cone coordinates these quantities read $W^{\pm\pm}=\delta S_{(m)}/\delta e^{\mp\mp}$, $W=\delta S_{(m)}/\delta X$ and $W_{\pm}=\delta S_{(m)}/\delta \psi^{\mp}$.
We add some comments regarding the physical meaning of the
eqs.~of motion: \eqref{eq:gPSMeom3.1} allows to eliminate the
auxiliary fields
$X^a$ as directional derivatives of the dilaton $X$;
\eqref{eq:gPSMeom3.2} contains the relevant information about the
Casimir function related to mass (see \eqref{eq:scs7a} below) and
corresponds to the ``Einstein equations'' in the sense that
minimally coupled matter may enter these eqs.~as a source
term $W^a$; \eqref{eq:gPSMeom3.3} is the fermionic counterpart of these
``Einstein equations''; \eqref{eq:neweom1} exhibits charge
conservation of the $U(1)$ gauge field $\Aapp$ and yields the second
Casimir (see \eqref{eq:scs7b} below); \eqref{eq:gPSMeom4.1} yields
curvature (matter coupled non-minimally to the dilaton may enter
with a source term $W$); \eqref{eq:gPSMeom4.2} is the torsion
condition; \eqref{eq:gPSMeom4.3} provides the equation for
gravitino propagation; \eqref{eq:neweom2} allows to express the
dilaton on-shell as the dual field strength related to $A$ plus a soul
contribution.

The Poisson tensor has at least two commuting Casimir functions that
can be chosen as
\begin{equation}
  \label{eq:scs7a}
  C = \half{1} X^a X_a - \inv{8} (X^2 - \B)^2 + \inv{8} X \chi^2\,,
\end{equation}
being related to the mass, and the charge
\begin{equation}
  \label{eq:scs7b}
  c = \B\,.
\end{equation}
We fix an eventual additive
constant in \eqref{eq:scs7a} by the requirement that $C=0$ on
states respecting both supersymmetries. This happens to be the
case in the definition \eqref{eq:scs7a}. Also, the
Casimir $C$ as defined by (\ref{eq:scs7a}) and $\mathcal{C}^{(g)}$
as defined in \cite{Grumiller:2003ad} differ by $\B^2/8$ from each
other. Finally, note that $C<0$ for a
positive mass.

Of course, in order to solve the bosonic part we set
$\chi^\alpha=0=\psi_\alpha$, thereby simplifying
\eqref{eq:gPSMeom3.1}-\eqref{eq:scs7b} considerably. Indeed, as we
will see in the next paragraph it becomes almost a triviality to
construct the classical solutions extending over basic
Eddington-Finkelstein patches.

\paragraph{Basic patches}

Let us start with an assumption: $X^{++}\neq 0$ for a given patch (the indices $++$ and $--$ refer to light cone components introduced in appendix \ref{app:A}.).\footnote{To get some physical intuition as to what this condition could mean: the quantities $X^a$, which are the Lagrange multipliers for torsion, can be expressed as directional derivatives of the dilaton field by virtue of \eqref{eq:gPSMeom3.1} (e.g.~in the second order formulation a term of the form $X^aX_a$ corresponds to $(\nabla X)^2$). For those who are familiar with the Newman-Penrose formalism: for spherically reduced gravity the quantities $X^a$ correspond to the expansion spin coefficients $\rho$ and $\rho^\prime$ (both are real).} If it vanishes a (Killing) horizon is encountered and one can repeat the calculation below with indices $++$ and $--$ swapped everywhere. If both vanish in an open region by virtue of \eqref{eq:gPSMeom3.1} a constant dilaton vacuum emerges, which will be addressed separately below. If both vanish on isolated points the Killing horizon bifurcates there and a more elaborate discussion is needed \cite{Klosch:1996qv}. The patch implied by this assumption is a ``basic Eddington Finkelstein patch'', i.e., a patch with a conformal diagram which, roughly speaking, extends over half of the bifurcate Killing horizon and exhibits a coordinate singularity on the other half.

In such a patch one may redefine $e^{++}=X^{++} Z$ with a new 1-form $Z$. Then \eqref{eq:gPSMeom3.1} implies $e^{--}=\extd X/X^{++}+X^{--}Z$ and the volume form reads $\epsilon=e^{++}\wedge e^{--}=Z\wedge \extd X$. The $++$ component of \eqref{eq:gPSMeom3.2} yields for the connection $\omega=-\extd X^{++}/X^{++}+ZV(X,Y)$. One of the torsion conditions \eqref{eq:gPSMeom4.2} then establishes $\extd Z=0$, i.e., $Z$ is closed. Locally, it is also exact: $Z=\extd u$. It is emphasized that, besides the two Casimir functions, this is the only integration needed! Thus, after these elementary steps one obtains already the line element
\begin{equation}
  \label{eq:lieelement}
  \extd s^2=2e^{++}e^{--}=2\extd u\,\extd X+2X^{++}X^{--}\extd u^2\,,
\end{equation}
which nicely demonstrates the power of the first order gravity/gPSM formalism.

It should be recalled that the set of classical solutions is
labelled by the two Casimir functions $C$ in \eqref{eq:scs7a},
corresponding to minus the mass, and $c$ in \eqref{eq:scs7b},
corresponding to the $U(1)$ charge. Exploiting their explicit form the bosonic
part\footnote{The fermionic part of the solutions can be discussed
along the lines of refs.~\cite{Ertl:2000si,Bergamin:2003am}.} of the line
element (in Eddington-Finkelstein gauge) may be presented as
\begin{align}
  \label{eq:cs6}
  \extd s^2&=2\diff u\,\diff X + K(X;C,c)\diff u^2\,, \\
  K(X;C,c)&=2C+(X^2-c)^2/4\,. \label{eq:killing}
\end{align}
Evidently there is always a Killing vector
$k\cdot\partial=\partial/\partial u$ with associated Killing norm
$k\cdot k=K(X;C,c)$. Now the geometrical meaning of the quantity
$X^{++}X^{--}$ is clear: if and only if this product vanishes a
(Killing) horizon is encountered. Thus, the condition $X^{++}\neq
0$ implies that the solution \eqref{eq:cs6}, \eqref{eq:killing} is
valid in a basic outgoing (ingoing) Eddington-Finkelstein patch,
i.e., it extends over outgoing (ingoing) horizons, only, as
opposed to \eqref{2metric} which does not extend over any kind of
horizon. Global solutions can be obtained by gluing together these
basic patches appropriately. Note that with the simple coordinate
transformation $\extd u= (2/c) (\extd t-\extd x/\killing)$, $\extd
X= (c/2) \extd x$ and the identification $\killing= (2/c)
\sqrt{K}$ the line element \eqref{eq:cs6} is equivalent to the one
in \eqref{2metric}. The coordinate singularities in that
transformation at $\killing=0=K$ should be observed. The (bosonic)
scalar curvature becomes\footnote{While \eqref{eq:cs6} with
\eqref{eq:killing} is not valid for constant dilaton vacua,
\eqref{eq:cs7} extends to these special solutions.}
\begin{equation}
  \label{eq:cs7}
  r = - \diff^2K/\diff X^2 = c-3X^2\,,
\end{equation}
in accordance with \eqref{rsol} (recall that on-shell $X=f$).
Obviously, solutions with constant curvature are only possible for the constant dilaton vacua discussed below.
Note that $r$ is independent of the ``mass'' $C$. This somewhat counter intuitive feature is a consequence of
the chosen conformal frame, in which generic solutions locally yield the same bosonic geometry as the ground
state.\footnote{This feature is well-known from generic dilaton gravity \cite{Grumiller:2002nm} if a special conformal frame is used. For instance, the 2D part of the Schwarzschild BH
\[
  (\extd s)^2 = 2\diff u\diff r + \left(1-\frac{2M}{r}\right)\diff u^2\,,
\]
by a conformal transformation can be mapped onto
\[
   (\extd s)^2 = 2\diff u\diff\tilde{r} + \left(\frac12 \sqrt{\tilde{r}}-2M\right)\diff u^2\,,
\]
which displays the same property as \eqref{eq:killing} regarding
the mass. Reverting this argument it is possible to transform
(\ref{eq:cs6}) to a conformal frame in which $C$ is a factor in
front of a curvature singularity, see Appendix \ref{app:C}.}

It is emphasized that \eqref{eq:killing} comprises {\em all} non-constant dilaton solutions, in particular the ones given above in \eqref{ds4} emerging as the special case $C=0$ and describing the patch $X\in(-\sqrt{c},\sqrt{c})$ (the explicit form of the coordinate transformation from \eqref{eq:killing} with $C=0$ to \eqref{ds4} can be found in sect. 2.2 of \cite{Grumiller:2003ad}).

\paragraph{Global properties}

In order to obtain the global Carter-Penrose diagrams one has to investigate the behavior of geodesics at the various boundaries. The basic idea is that if a, say, null geodesic reaches with finite affine parameter a boundary which does not exhibit a curvature singularity then spacetime may be extended. It is not the intention of the present work to repeat this discussion; rather, the reader is referred to \cite{Grumiller:2003ad}.

\subsubsection{Switching on SUSY}\label{se:1783}

The classification of SUSY solutions is based upon the discussion
in ref.~\cite{Bergamin:2003mh}. It exhausts all smooth solutions
preserving (at least) half of the SUSYs. A relevant observation is
that the solutions that preserve some SUSY are equivalent to those with a
vanishing body of the Casimir function $C$. This
result can be anticipated on general grounds as the body of $C$ is proportional to the
determinant of the fermionic part of the Poisson tensor \eqref{eq:scs4}. An
interesting non-smooth solution, the BPS-kink, will be discussed as well.

\paragraph{Constant dilaton vacua}

There is a very special class of solutions\footnote{From the gPSM point of view these solutions are remarkable as they allow not only for two but for four bosonic Casimir functions, i.e., the bosonic part of the Poisson tensor has minimal rank.} called ``constant dilaton vacua'', i.e., $X=\rm const.$ globally.
For the current model these are the two
$\mathbb{Z}_2$ symmetry breaking solutions \eqref{fsol} (B) and the symmetry
preserving one \eqref{fsol} (A), all of them yielding constant curvature (in D=2
and also in D=3). Provided the fermions are set to zero the
$\mathbb{Z}_2$ symmetry breaking solutions indeed preserve half of the
SUSYs because they are ground state geometries (in the sense that
$C=0$), in accordance with our previous discussion below Eq.~\eqref{paralell2}.
Curvature and dual field strength are given by (cf.~case (B) in \eqref{fsol}, \eqref{rsol})
\begin{equation}
  \label{eq:scs1783}
  r=-2c\,,\quad f=\pm\sqrt{c}\,.
\end{equation}
The $\mathbb{Z}_2$ symmetry preserving one violates both SUSYs ($r=+c$, $f=0$). The fact that symmetry breaking solutions preserve part of SUSY while symmetry preserving ones break it may have been expected on general grounds.

\paragraph{BPS states with non-constant dilaton}

In general, there exist two classes of BPS states:

{\it Non-vanishing fermion fields.} If the fermionic background is
non-trivial for a BPS state the bosonic one has to be trivial,
viz., Minkowski space \cite{Bergamin:2003mh}. However, none of the solutions
\eqref{eq:cs6} describe Minkowski space. Thus, there can be no BPS
states with non-vanishing fermions.

{\it Vanishing fermion fields.} In this case only \emph{the
variation of the fermions} is important, as any variation of a
boson is proportional to some fermion field that vanishes by
assumption. All solutions with $C=0$ are found to be BPS states
provided the symmetry parameters are related in a specific way.
The Killing norm is given by the complete square
\begin{equation}
  \label{eq:killingbps}
  K_{BPS}(X;c)=(X^2-c)^2/4=u^2/4\geq 0\,.
\end{equation}
The expected \cite{Gibbons:1981ja}
positivity result is recovered implying that all horizons have to
be extremal.  Curvature\footnote{Note that the curvature
singularities are null complete but incomplete with respect to
non-null geodesics. Thus, you cannot send an SMS with your mobile
phone to it, but you can write it on a piece of paper and throw it
there. This somewhat counter intuitive feature is a consequence of
the chosen conformal frame.} and dual field strength read
\begin{equation}
  \label{eq:bpsrf}
  r=c-3X^2\,,\quad f=X\,,
\end{equation}
depending on the non-Killing coordinate $X\in(-\infty,\infty)$.
The global structure has been discussed in
\cite{Grumiller:2003ad}; keeping that notation\footnote{Cf.\ however the
  discussion below eq.\ \eqref{eq:scs7b}. In fig.~2 of that work BPS restricts to the curve starting
from the origin of that diagram.} we recall the two Carter-Penrose
diagrams with $C=0$. Bold lines denote boundaries (in the present case
they are curvature singularities),
dashed lines Killing horizons (all of them are extremal here) and
ordinary curved lines are hypersurfaces of constant $f$ (see fig.~\ref{fig:CPs}).

\FIGURE[t]{
\begin{minipage}[c]{0.1\linewidth}
{\bf B1b:}
\end{minipage}
\begin{minipage}[c]{0.25\linewidth}
\epsfig{file=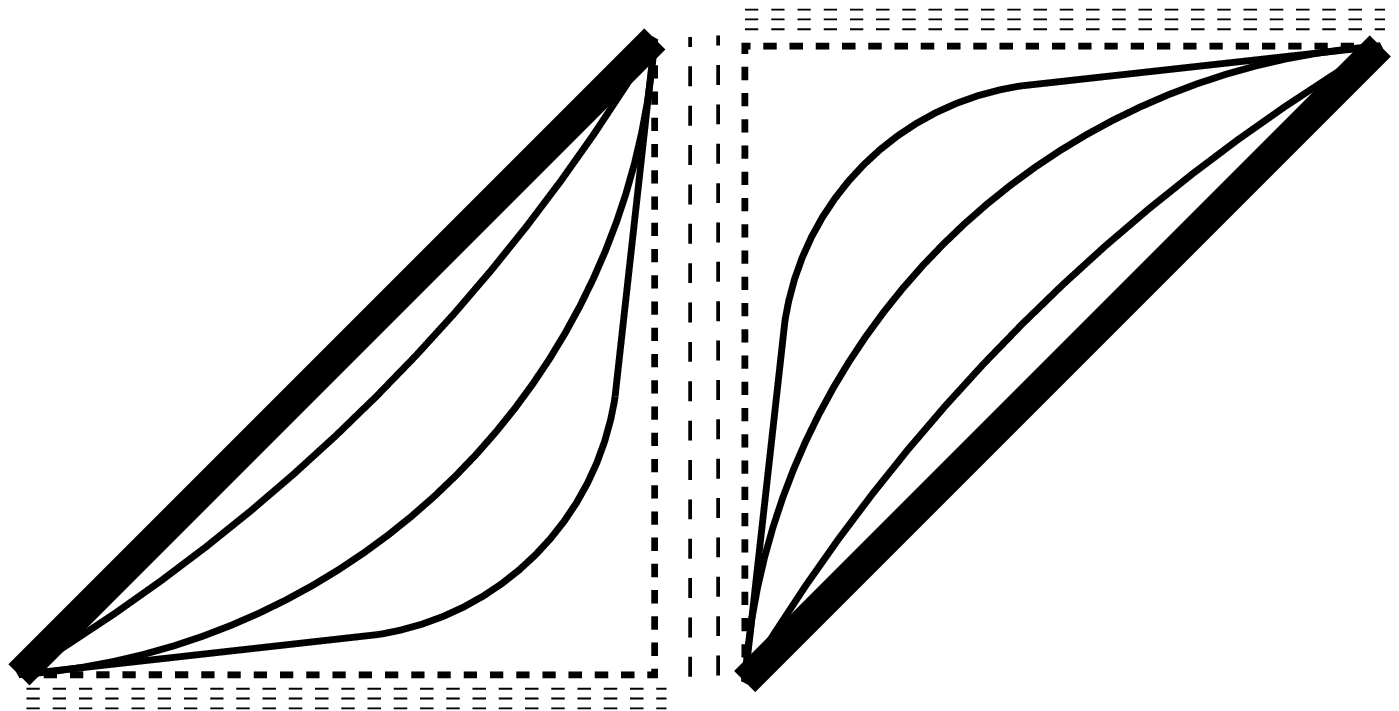,width=\linewidth}
\end{minipage}
\begin{minipage}[c]{0.1\linewidth}
\,
\end{minipage}
\begin{minipage}[c]{0.1\linewidth}
{\bf B2b:}
\end{minipage}
\begin{minipage}[c]{0.35\linewidth}
\epsfig{file=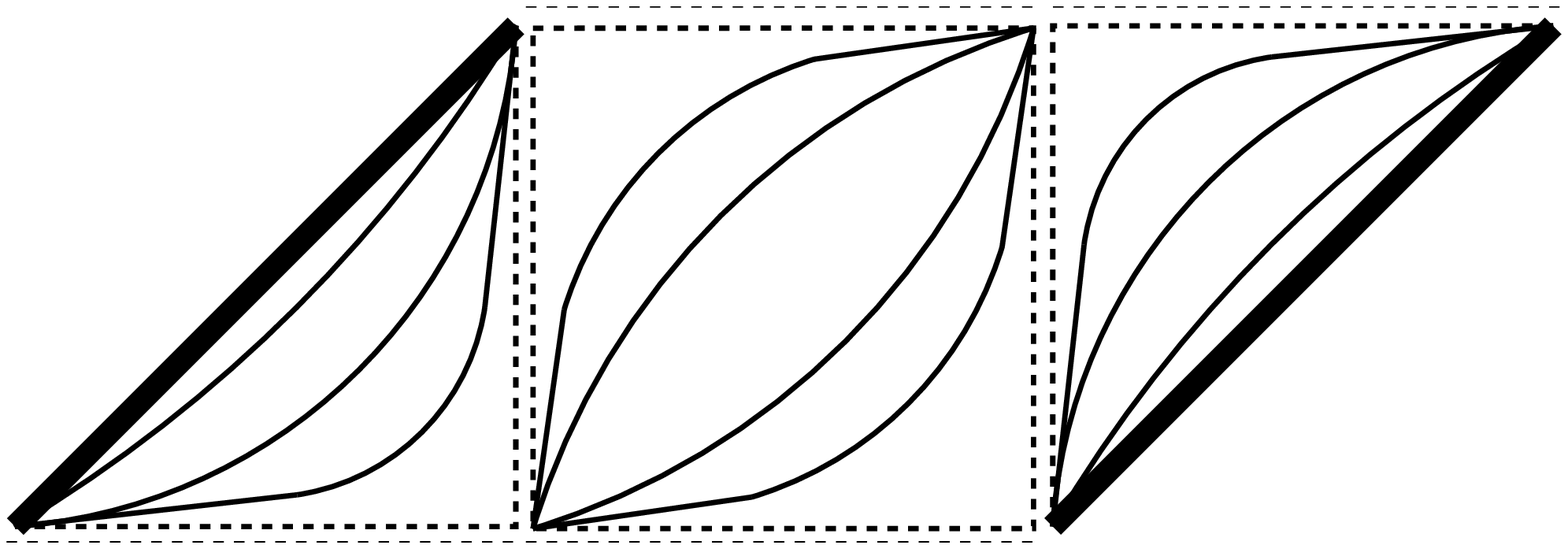,width=\linewidth}
\end{minipage}
\caption{Basic Carter-Penrose patches for BPS solutions with non-constant dilaton}
\label{fig:CPs}}

The former {\bf B1b} arises as a special case of the latter {\bf B2b} when the
``kink''-region (the square patch in the middle) is shrunk to zero
size by requiring $c=0$. Note that both diagrams have to be
rotated counter clockwise by $45^o$ if ``time'' is plotted
vertically and ``space'' horizontally.\footnote{In
ref.~\cite{Grumiller:2003ad} in Section 3.1 in the third paragraph
the words ``time'' and ``space'' have to be exchanged.
Alternatively, of course one could rotate all Carter-Penrose
diagrams in that work by $90^o$.} They can be extended to an
infinite strip, much like the Carter-Penrose diagram of the
Reissner-Nordstr\"om black hole (RN BH). Of course, one can identify
periodically to get a finite strip; M\"obius-strip like
identifications are possible as well.

\subsection{The BPS-kink}\label{se:2}

There is an interesting non-smooth candidate for a BPS state: the
kink solution of \cite{Guralnik:2003we} which consists of the
square patch of {\bf B2b} patched continuously to the two symmetry
breaking solutions $f=\pm \sqrt{c}$, $r=-2c$. On general grounds
\cite{Israel:1966rt} such a patching induces a matter flux along
the hypersurface of patching. This was shown explicitly in
\cite{Grumiller:2003ad} for the case under consideration, and it
was found that the induced matter flux has to be (anti-)self
dual.\footnote{Whether self- or anti-self dual depends on the
patching: one can glue the two symmetry breaking solutions either
on the left and right horizon in {\bf B2b} or on the upper and lower
one. We will treat only self dual fluxes explicitly but the anti-self dual ones may be obtained by exchanging $+$ indices with $-$ indices in all subsequent considerations.}

We briefly recall that discussion and translate it into the second
order formulation in order to pinpoint the source of
non-smoothness in the oxidized theory in 3D: The dual field
strength $f$ as a function of the non Killing coordinate remains
continuous but is not differentiable anymore: in the kink region
it is linear according to (\ref{eq:bpsrf}), while in the two
symmetry breaking vacua it is constant (\ref{eq:scs1783}). Because
the condition of vanishing Cotton tensor $C^{\mu\nu}=0$
(cf.~\eqref{cottontensor}) implies a second order differential
equation for $f$ a $\delta$-function appears in it generating the
induced matter flux. Thus, the eqs.~of motion have to be
modified to $C^{\mu\nu}=T^{\mu\nu}$, where $T^{\mu\nu}$ is the
induced matter flux on the hypersurface of patching. Note that the
non-smoothness is difficult to spot in the coordinate system used
in Eq.~\eqref{ds4} because the patching occurs in the ``asymptotic
region'' $x=\pm\infty$. However, as the affine distance to this
region is finite the solution can be extended and the smooth
extension in 2D is nothing but {\bf B2b}.

In general such a patching procedure might destroy the BPS
property. However, for the kink solution there are two reasons for
optimism: 1.~all patches preserve half of the SUSYs and
2.~the induced matter flux is (anti-)self dual. In
addition it is purely bosonic as the fermionic background is
trivial and hence no fermionic matter flux is induced. Although
each of these properties alone is not sufficient to guarantee BPS,
taken together they are. Technically, the crucial observation is
that the conservation equation in presence of matter, (7.4) of
\cite{Bergamin:2003mh}, reduces to the one in the absence of
matter because 1.~no fermions are present, 2.~the matter component
$X^{++}W^{--}$ vanishes as the matter component $W^{--}=0$ and 3.~the remaining matter contribution $X^{--}W^{++}$ vanishes as well because $W^{++}\neq 0$ is valid only on the horizon where $X^{--}=0$. Thus, $C=0$ is still the ground state
which is necessary\footnote{That it is also sufficient in the
present case can be checked easily by considering the restrictions
on the SUSY transformation parameters $\epsilon_\pm$ in sect.~4 of
\cite{Bergamin:2003mh}. Note that within each patch $C=0$ is both necessary and sufficient for BPS provided the fermions vanish, but one needs to check in addition the compatibility of patching.} for BPS states with
vanishing fermions.

\newcommand{\xp}{x_+}
\newcommand{\xm}{x_-}
\newcommand{\xk}{\theta_{\rm kink}}
Introducing the coordinate $x\in(-\infty,\infty)$ which coincides
with $X$ in the kink region $x\in[-\sqrt{c},\sqrt{c}]$, the
following situation is encountered globally (recalling that the
horizons are located at $x=\pm\sqrt{c}$ and defining
$x_\pm:=x\pm\sqrt{c}$, $\xk:= \theta(\xp)-\theta(\xm)$ and
$\theta(0):=1/2$)
\begin{align}
\epsilon_+ = \epsilon_+(x) \,&\xk\,,  \\
\epsilon_- = \epsilon_-(x) \,&\xk + \epsilon_-^{(l)} \theta(-\xp) + \epsilon_-^{(r)} \theta(\xm) \,,\\
f = x \,&\xk - \sqrt{c}\left(\theta(-\xp)-\theta(\xm)\right) \,,\\
r = (c-3x^2) &\underbrace{\xk}_{\rm kink} -2c \bigl(\underbrace{\theta(-\xp)}_{\rm left\,\,AdS}
+ \underbrace{\theta(\xm)}_{\rm right\,\,AdS}\bigr)\,,
\end{align}
with the following continuity properties of the symmetry parameters
\begin{align}
\epsilon_+(\pm\sqrt{c})&=0\,,\\
\epsilon_-(\sqrt{c})   &=\epsilon_-^{(r)}\,,\\
\epsilon_-(-\sqrt{c})  &=\epsilon_-^{(l)}\,.
\end{align}
Additionally, the only non-vanishing component of the
energy-momentum tensor induced by patching is given by
\cite{Grumiller:2003ad}
\begin{equation}
  \label{eq:emom}
  T^{uu}=\delta(\xm)-\delta(\xp)\,.
\end{equation}
where $u$ is the lightlike coordinate used in (\ref{eq:cs6}). In a
sense, these matter fluxes localized on the extremal horizons can
be considered as stabilizing the kink.

\FIGURE[t]{\parbox{\linewidth}{\centering \epsfig{file=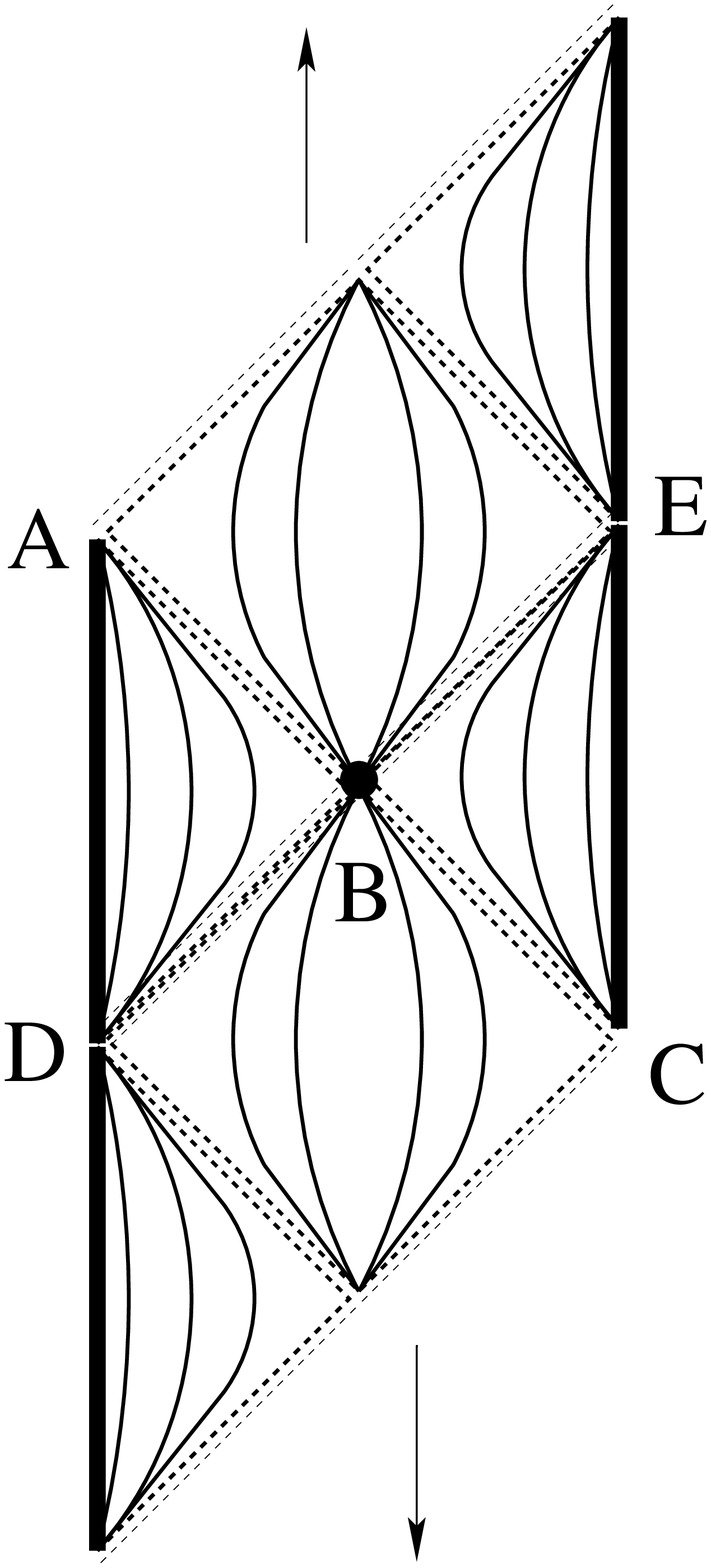,width=0.2\linewidth}
\caption{Maximally extended Carter-Penrose diagram for the
BPS-kink.}} \label{fig:strip}}
In Fig.~\ref{fig:strip} the Carter-Penrose diagram for the kink is
depicted. Note that the boundaries do not represent a curvature
singularity. Time is plotted vertically.
The (anti-)self dual shock waves propagate from A (D) to B and from
C (E) to B, respectively. The point B is the bifurcation point. The diagram can be
continued above and below. It has strip-like topology with
singularities at the boundaries. The outer triangle patches are
the $AdS$ vacua, the inner square patch surrounded by extremal
Killing horizons is the kink region.

\paragraph{Physical instances of the kink} As the model under consideration has no propagating physical modes it is a fair question whether the kink solution is a gauge artifact or a physical entity. At first glance there seems to be no physics, as \eqref{ds4} by a {\em regular} conformal transformation can be mapped to flat space.
On the other hand, one can argue on general grounds that the kink solution is as ``real'' as the (extremal) RN BH, because the latter may be described as well by a (g)PSM with four target space coordinates and a degenerate (bosonic) Poisson tensor of rank 2, the two Casimir functions being related to ADM mass and total charge. In this context one should be rather careful with the definition of mass (for a recent discussion in the framework of 2D dilaton gravity cf.~Appendix A of \cite{Grumiller:2004wi}).
In particular, the conformal frame plays an important role here. The purpose of the rest of this section is to elucidate these issues and to clarify which of the features of the kink solution actually are physical.

In the chosen conformal frame all kink solutions have the same energy, namely the energy of the ground state ($C=0$). They differ by the value of the $U(1)$ charge $c$, but qualitatively they are essentially the same (they yield the same Carter-Penrose diagram) apart from the special case $c=0$. Actually, this is again equivalent to the corresponding properties of the extremal RN BH: all of them have the same BPS mass\footnote{For the RN BH the BPS mass is defined as $M_{BPS}:=M_{ADM}-|Q|$. Thus, by definition all BPS states have vanishing BPS mass, while their ADM mass is given by the charge. Our definition of the Casimir function is adjusted such that $C$ vanishes for BPS states.\label{fn:BPS}} and the family of extremal solutions may be labelled by the value of the $U(1)$ charge.
Note, however, that the charge $c$ is not quantized and thus there is a continuous spectrum of kinks with the same energy.

It is recalled that the 3D scalar field has been set to 1 and thus it is regular everywhere. Therefore, the only singularities in this conformal frame are the ones induced by the metric or the gauge field. As has been shown above, a curvature singularity exists only for $X\to\pm\infty$, which is also the locus where the gauge field diverges. Thus, eventual conformal transformations should be regular everywhere except in the asymptotic regions $X\to\pm\infty$ where they may be singular, in order to avoid the introduction of spurious singularities.

Why is it of interest to study different conformal frames at all? First of all, as the theory exhibits conformal invariance it may be considered as a consistency check. Additionally, any mass definition depends crucially on the selected ground state geometry and the latter may exhibit slightly different behavior in different conformal frames. Moreover, it may well be that some features of the kink encountered above are actually an artifact of the chosen frame and thus it is of interest to separate these from the true physical features.

Appendix \ref{app:C} is devoted to a study of conformal transformations with a conformal factor monomial in $X$. However, all these transformations destroy the kink solution because they are singular at $X=0$ and thus either a curvature singularity is introduced at $X=0$ (so there is no interpolating solution available between positive and negative $X$ anymore) or the point $X=0$ is mapped to infinity (and thus the kink solution may not pass through this point). The singularity of the conformal factor at $X=0$ is in contradiction with our requirement of regularity and thus one needs at least binomials. Rather than repeating a (somewhat tedious) study analogous to appendix \ref{app:C} a short cut will be taken: having in mind the nice features of the $AdS$ ground state \eqref{eq:AdSgs} and the requirement of regularity at $X=0$ it is easy to find a convenient conformal factor
\begin{equation}
  \label{eq:cf}
  e^{Q} = \alpha X^2 + \delta\,,
\end{equation}
with some positive $\alpha,\delta$. Obviously, \eqref{eq:cf} is regular everywhere except for the asymptotic region in accordance with our requirements.
The constant $\delta$ defines how geometry is rescaled at the origin
$X=0$ and may be set to 1, while $\alpha$ sets the scale for the
asymptotically constant curvature \eqref{eq:AdSgsr}. With
$\tilde{X}=\alpha X^3/3 + \delta X$ the line element is transformed
to ($X$ is understood to be a function of $\tilde{X}$)\footnote{Explicitly: $X=W^{-1/3}\delta-W^{1/3}/\alpha$ with $W=\frac32 \alpha^2\tilde{X} \Big(\sqrt{1+4\delta^3/(9\alpha\tilde{X}^2)}-1\Big)$}
\begin{equation}
  \label{eq:thisisaverylonglabel}
  \extd s^2=2\extd u\extd \tilde{X} + (\alpha X^2+\delta)(2C+(X^2-c)^2/4)\extd u^2\,.
\end{equation}
Note that \eqref{eq:cf} by virtue of \eqref{eq:Z} implies the somewhat unusual bosonic potential $Z=2\alpha X/(\alpha X^2+\delta^2)$ which only asymptotically has the ``standard form'' $Z=\beta/X$ with $\beta=2$.
The bosonic second order action in this frame instead of \eqref{csd=2} takes the form of \eqref{csd=2phi} with $\phi=e^Q=\alpha X^2+\delta$. The relation between the second order gauge field $\A$ and the first order one $\Aapp$ up to pure gauge terms now reads
\begin{equation}
  \label{eq:gf12}
  \A=\frac{X^2}{\int^Xe^{Q(X^\prime)}X^\prime\extd X^\prime} \Aapp = \frac{X^2}{2}\extd u\,.
\end{equation}
By construction $\A$ is invariant under conformal transformations while $\Aapp$ is not. Correspondingly, the dual field strength is related to the dilaton field by
\begin{equation}
  \label{eq:fX}
  f=-Xe^{-Q(X)}=-\frac{X}{\alpha X^2+\delta}\,.
\end{equation}
One can solve uniquely for $X$ in terms of $f$ noticing that $f\to 0$ implies $X\to 0$.

In the conformal frame implied by \eqref{eq:cf} geometry approaches asymptotically $AdS$ and thus it may be joined continuously to the symmetry breaking constant dilaton vacua if $\alpha$ is chosen as $\alpha=9/(4c)$. Therefore, the conformal factor depends on the charge $c$ but is independent from the mass $C$. No patching at the Killing horizons is necessary and consequently no matter fluxes are induced. The solution avoids curvature singularities everywhere. Up to globally regular conformal transformations this conformal frame is unique. Thus, any physical discussion should be based upon this frame.

To summarize, the physical features of the kink solution are 1.~the ground state property (they are all solutions of lowest energy), 2.~the nonvanishing $U(1)$ charge $c$, 3.~the existence of two extremal Killing horizons in a basic Eddington-Finkelstein patch, 4.~the fact that it interpolates between the two symmetry breaking $AdS$ vacua. In the original frame the last point could not be achieved smoothly but only in the presence of (self-dual) matter fluxes. In the frame implied by the conformal factor \eqref{eq:cf} all solutions asymptotically approach $AdS_2$ and thus the symmetry breaking $AdS$ vacua may be patched continuously and without inducing matter fluxes provided $\alpha$ is tuned to $9/(4c)$. We will comment on this asymptotic behavior in more detail in sect.~\ref{se:AAdS}.

\subsection{Hawking temperature}\label{se:ht}

As some of the solutions exhibit (Killing) horizons the Hawking effect \cite{Hawking:1975sw} is of relevance. There are many ways to calculate the Hawking temperature, some of them involving the coupling to matter fields, some of them being purely geometrical. Because of its simplicity and since we do not intend to introduce matter explicitly, we will restrict ourselves to a calculation of the geometric Hawking temperature as derived from surface gravity (cf.~e.g.~\cite{waldgeneral}).\footnote{If defined in this way Hawking temperature turns out to be independent of the conformal frame. Thus, we may stick to the simple conformal frame where no kinetic term of the dilaton arises (cf.~Eq.~\eqref{eq:ctsg} in Appendix \ref{app:C}). For a review on Hawking radiation in D=2 cf.~e.g.~\cite{Kummer:1999zy}.}
The latter can be calculated by taking the normal derivative $\extd/\extd X$ of the Killing norm $K(X;C,c)$ given in Eq.~\eqref{eq:killing}, evaluated on one of the Killing horizons $X=X_h$, where $X_h$ is a solution of $K(X_h;C,c)=0$:
\begin{equation}
  \label{eq:ht}
  T_H = \frac{1}{4\pi} \Big|\frac{\extd}{\extd X}K(X;C,c)\Big|_{X=X_h} =  \frac{1}{2\pi} \Big|V(X,Y) \Big|_{X=X_h}^{Y=c} = \frac{1}{4\pi} \left|X_h(c-X_h^2)\right|\,.
\end{equation}
The numerical prefactor in \eqref{eq:ht} can be changed by a redefinition of the Boltzmann constant. It has been chosen in accordance with refs.~\cite{Kummer:1999zy,Grumiller:2002nm}.
The zeros of the Killing norm are given by
\begin{equation}
  \label{eq:killing0}
  X_h = \pm \sqrt{c\pm2\sqrt{2M}}\,.
\end{equation}
Here it should be recalled that for sake of backward compatibility $C\leq 0$ for positive mass configurations \cite{Grumiller:2002nm}; we have thus defined $M=-C$, thereby fixing an irrelevant scale factor between minus the Casimir and mass to unity. Henceforth $M\geq 0$ will be assumed; the bound is saturated only for the ground state geometries. The Hawking temperature may be presented as
\begin{equation}
  \label{eq:ht2}
  T_H = \frac{1}{2\pi}\sqrt{2M}\sqrt{c\pm2\sqrt{2M}} \,,
\end{equation}
where the $+$ sign refers to the outer horizons (which exist as long as $c\geq-2\sqrt{2M}$) and the $-$ sign to eventual inner ones (which may exist only for $c\geq2\sqrt{2M}$). Thus, in the large $M$ limit one obtains $T_H\propto M^{3/4}$, while in the large $c$ limit one gets $T_H\propto\sqrt{cM}$. Remarkably, the specific heat
\begin{equation}
  \label{eq:specificheat}
  C_s=\frac{\extd M}{\extd T_H}=\frac{4\pi^2 T_H}{c\pm 3\sqrt{2M}}
\end{equation}
 is always positive on outer horizons. It may become negative on inner
 horizons provided that $3\sqrt{2M}>c>2\sqrt{2M}$. Somewhat
 unexpectedly for $c=3\sqrt{2M}$ the inverse specific heat vanishes on
 the inner horizon. In order to compare these results with similar
 well-known ones, it is recalled that for the Schwarzschild BH the
 specific heat is always negative (cf.~e.g.~\cite{waldgeneral}), while
 for the Witten BH \cite{Mandal:1991tz} the inverse specific heat
 vanishes. It is worthwhile mentioning that the Hawking-Page
 transition between an $AdS$-Schwarzschild BH and pure $AdS$ filled
 with radiation \cite{Hawking:1982dh} occurs for $C_s^{-1}=0$. Thus,
 although the thermodynamical analysis of a multi-horizon geometry
 clearly is more involved, a similar behavior might be expected in our
 case at $c=3\sqrt{2M}$, corresponding to a critical temperature of
\begin{equation}
T_{\rm critical}=\frac{1}{2\pi}(c/3)^{3/2}\ .
\end{equation}
In addition to being interesting on its own, the result \eqref{eq:ht2} provides a consistency check for our BPS solutions: according to a well-known argument due to Gibbons \cite{Gibbons:1981ja} BPS solutions must have vanishing Hawking temperature. Indeed, for $M=0$ it vanishes. Incidentally, for $M=c^2/8$ and $c>0$ there is also a non-BPS solution which nevertheless has vanishing Hawking temperature. It corresponds to the temperature dictated by the extremal inner horizon in the solution {\bf B3} of \cite{Grumiller:2003ad} which additionally exhibits two nonextremal outer horizons  and thus cannot be BPS globally.

It could be of interest to check our straightforward analysis by one of the more involved derivations of the Hawking effect, in particular one where matter is present.

\section{Higher dimensional perspective}
In this section, we would like to analyze the SUSY of the 2D kink
solution, but from a 3D viewpoint. Finally, we will also provide a
connection with M-theory membranes, that was basically spelled out
in \cite{Cacciatori:2004rt}. We will also comment on $AdS$/CFT
aspects of the 11D SUGRA solution.

Let us start by studying the kink geometry in 3D SUGRA. In this
case the line element is \eqn{kink}{\extd s^2= e^{2
g(x)}(-\extd x^2 - \extd r^2 - 2\killing(x) \extd t \extd r)\,.}
The coordinates $t,x$ are equivalent to the corresponding ones in D=2,
while $r$ denotes the Killing direction that has been eliminated
by Kaluza-Klein reduction. We take this form of the conformal
factor, because we are interested in solutions that are basically
2D and because we will consider static solutions. The presence of
the conformal factor $e^{2 g(x)}$ will be important in the
following. With this metric, we can associate a set of Vielbeine
\eqn{vilekinkc}{e^{0}=e^{g(x)}\killing(x) \extd t, \;\;
e^{1}=e^{g(x)} \extd x, \;\;
 e^{2}= e^{g(x)}(\extd r + \killing(x) \extd t), }
which yields the spin connection $\omega^{AB}_\mu$
\eqn{spinkinkc}{\omega_{r}^{01}=  \omega_{x}^{02}=-\frac{\killing'(x)}{2\killing(x)},\;\; \omega_{t}^{01}=
\omega_{t}^{21}= \frac{\killing'(x)}{2}+\killing(x) g'(x) ,\;\; \omega_r^{21}= g'. }
Now, it will be assumed that the spinor $\epsilon$
depends only on the coordinate $x$. Consequently, the SUSY transformations \cite{Deser:1983sw}
\begin{equation}
  \label{eq:susydeserkay}
  \delta\psi_\mu=
  \partial_\mu \epsilon+ \frac{1}{2}\omega_{\mu}^{AB}\sigma_{AB}\epsilon
\end{equation}
with $\sigma_{AB}:=\Gamma_A\Gamma_B$ imply
\eqn{vartc}{\delta\psi_t=0\to \partial_t\epsilon
+(\frac{\killing'}{2}+ \killing g') (\sigma_{01}
+\sigma_{21})\epsilon=0} \eqn{varyc}{\delta\psi_x=0\to
\partial_x\epsilon -\frac{\killing'}{2\killing} \sigma_{02}
\epsilon=0} \eqn{varcr}{\delta\psi_r=0\to \partial_r\epsilon
+(\frac{\killing'}{2\killing} \sigma_{10} +g'\sigma_{21})
\epsilon=0} To impose that (part of) the SUSY is preserved, then
we have  to solve, \eqn{e3c}{(\frac{\killing'}{2} + \killing
g')\sigma_{01} (1 + \sigma_{20}) \epsilon=0}
\eqn{e2c}{\partial_x\epsilon -  \frac{\killing'}{2\killing}
\sigma_{02}\epsilon=0} \eqn{e1c}{(
-\frac{\killing'}{2\killing}\sigma_{02}   + g'
\sigma_{21})\epsilon=0.} We can solve (\ref{e3c}) if\footnote{Note
that \eqref{proyc} is a direct consequence of \eqref{e3c}, unless
\eqref{ultima} holds. Since this is actually the case we
nevertheless {\em impose} \eqref{proyc} as an independent
constraint in order to be able to solve \eqref{e2c}.}
\eqn{proyc}{\sigma_{20}\epsilon=-\epsilon,} plugging into
(\ref{e2c}) gives \eqn{solespinor}{\epsilon(x)=e^{-\int
\frac{\killing'}{2\killing} \extd x}  \epsilon(0). } and in
solving (\ref{e1c}) \eqn{ultima}{g'=-
\frac{\killing'}{2\killing}.} So, this provides an eq.~for the
conformal factor, in terms of the function $\killing(x)$.

There are important points to be noticed here. On the one hand,
the fact that the conformal factor $g(x)$ is nonzero, allows for a
fraction of SUSY to be preserved. Indeed, had we taken a $g(x)=0$,
this would be reflected in the spin connection (\ref{spinkinkc})
and eqs.~(\ref{e3c}), (\ref{e1c}) would not have solution. Second,
is that the conformal factor is determined in terms of the
function $\killing(x)$. Third, that the form of the spinor and the
eq.~(\ref{vartc}) are analogous with eqs.~(\ref{gravi1}),
(\ref{gravi2}) in the 2D analysis.

The reader might wonder why (in a conformally invariant theory)
the conformal factor is determined by eq.~(\ref{ultima}). Indeed,
one may think that $any$ conformal factor will be the same. But
preservation of some fraction of SUSY imposes a constraint.
Indeed, while the projection (\ref{proyc}) is breaking part of the
SUSY, it is also spoiling the superconformal invariance, thus
leading to a unique conformal factor. Summarizing, all possible
conformal factors will be solutions of the eqs.~of motion
$C^{\mu\nu}=0$, but only a subset of those solutions is
supersymmetric. This subset is described (if our trial line
element is of the `kink' form) by metrics like
\eqn{kinksuperymm}{\extd s^2= \frac{1}{\killing(x)}(- \extd x^2 -
2 \killing(x) \extd t \extd r - \extd r^2 ) \;,} with scalar
curvature vanishing for arbitrary $A$. Notice also, that it is
possible that more general supersymmetric solutions exist, where
the conformal factor depends on more than one coordinate, but
still preserving the same trial for the line element. Finally, it
should be stressed that the conformal factor in
\eqref{kinksuperymm} becomes singular at Killing horizons of the
2D geometry $\killing(x)=0$. Thus, \eqref{kinksuperymm}
essentially is trivial and may be mapped to Minkowski space by a
globally {\em regular} conformal transformation in 2D, where
``gobally'' refers to a region where $A\neq 0$. Therefore, in 3D
the only SUSY solution is globally conformal to Minkowski space.

For sake of completeness we provide the general bosonic result for the 3D metric. It follows directly from the {\em Ansatz} \eqref{KK1} with the gauge field, expressed in a convenient gauge,\footnote{In other gauges the constant $c$ may be replaced by an arbitrary function of $u$ and an additional exact term $f(X)\extd X$ may be present in $a$.}
\begin{equation}
  \label{eq:gaugefield}
  \A = \frac{X^2-c}{2}\extd u\,.
\end{equation}
and the 2D line element \eqref{eq:cs6}, \eqref{eq:killing}:
\begin{equation}
  \label{eq:3Dmetric}
  \extd s^2= 2C\extd u^2+2\extd u \extd X - (X^2-c)\extd u\extd r -\extd r^2
\end{equation}
Only for BPS states the $\extd u^2$ term vanishes. The appearance
of the prepotential \eqref{eq:scs1} in \eqref{eq:gaugefield} and
\eqref{eq:3Dmetric} should be noted. The 3D scalar curvature
\begin{equation}
  \label{eq:curvature3D}
  R = c - \frac{5X^2}{2}\,,
\end{equation}
is independent from $C$ in this particular conformal frame. For consistency it can be checked easily that the Cotton tensor derived from \eqref{eq:3Dmetric} indeed vanishes. A coordinate transformation to conformal gauge is possible in principle but somewhat tedious; necessarily it has singularities at values of the dilaton field given by \eqref{eq:killing0}. Therefore, in 3D, much like in 2D, conformal gauge need not be a convenient choice for calculations and especially not for the discussion of the global structure. For the simple case $C=0$ by a gauge transformation $\A\to\A+2\extd X/(X^2-c)$ and a coordinate redefinition $\extd t=-c\extd u/2 - 2c\extd X/(X^2-c)^2$, $\extd x=-2\extd X/(X^2-c)$ the line element for $X\in(-\sqrt{c},\sqrt{c})$ may be presented in a way coinciding with (4.56) of \cite{Guralnik:2003we},
\begin{equation}
  \label{eq:GIJPgauge}
  \left.\extd s^2\right|_{C=0} = -\extd x^2-\frac{2\extd t \extd r}{\cosh^2{\frac{\sqrt{c}x}{2}}} - \extd r^2\,.
\end{equation}
An explicit transformation of \eqref{eq:GIJPgauge} to conformal gauge can be found in ref.~\cite{Jackiw:2004is}. 
We will comment in detail on the asymptotic behavior in sect.~\ref{se:AAdS}.

\subsection{M theory Perspective}
Let us now consider the same solution from a higher dimensional
perspective. We will try to summarize a part of the paper
\cite{Cacciatori:2004rt}, that deals with a system in 4D $U(1)$
gauged SUGRA that has the same eqs.~of motion that the kink
solves. Then, we offer an M-theory interpretation of the kink by
using the lifting prescription given in \cite{Cvetic:1999xp}. The
general idea is that a particular configuration in 4D SUGRA, gives
place to a set of eqs.~of motion that are exactly the ones derived
from our 2D action (\ref{eq:scs13main}). As is well known, the bosonic
part of \eqref{eq:scs13main} can be derived from reduction of the
action \eqref{cs} as is shown in ref.\ \cite{Guralnik:2003we}. So,
in this section we want to connect this three systems. Besides,
the outcome of \cite{Cacciatori:2004rt} is that the configuration
is supersymmetric, the same we have shown to happen in our 2D
system and from a 3D perspective.

Let us start by considering ${\cal N}=2$ 4D $U(1)$ gauged SUGRA;
the bosonic part of the Lagrangian reads
\eqn{4dgauged}{L=\sqrt{g}[ R- F_{\mu\nu}^2 +\frac{6}{l^2}].}

The field content of this SUGRA is given by the metric, a $U(1)$
gauge field $\Afour_\mu$ and a complex gravitino $\psi_\mu$. The
supersymmetry transformations look

\eqn{transf}{\delta e_\mu^a ={\rm Re}\,[\bar{\epsilon}\gamma^a\psi_\mu], \;\;\delta
\Afour_\mu={\rm Im}\,[\bar{\epsilon}\psi_\mu],\;\; \delta \psi_\mu=\hat{D}_\mu \epsilon,}
where
\eqn{der}{\hat{D}_\mu \epsilon= [\nabla_\mu(\omega) -\frac{i}{l}\Afour_\mu +\frac{1}{2l}\gamma_\mu
+\frac{i}{4}F_{ab}\gamma^{ab}\gamma_\mu]\epsilon}
and $\gamma_{ab}$ is the usual antisymmetrized product of gamma matrices.

In \cite{Cacciatori:2004rt} it was shown that a certain class of BPS solutions
can be represented as (in this section we use the signature $(-1,1,1,1)$ in
accordance with \cite{Cacciatori:2004rt}),
\eqn{metrica}{\extd s^2= - f(r,x)^2(\extd t-a(r,x) \extd z)^2 +\frac{1}{f(r,x)^2}\extd r^2 +
\frac{g(r,x)^2}{1+\delta(x)}\Big(\frac{(1+\delta(x))^2\extd z^2}{\cosh^4(x/2)} + \extd x^2\Big)\,,   }
with gauge field,
\eqn{gaugefield}{\Afour= s(r,x)\bigl((r^2 + \inv{4}) \extd t+
  \frac{\delta_0}{4} \extd z\bigr)\ .}
The coordinates $t,x,r$ essentially correspond to the 3D coordinates used in
\eqref{kink} and
\ba
&  f(r,x) = \frac{r^2 + 1/4}{g(r,x)}\,,\quad & a(r,x) = \frac{1+\delta(x)}{4(r^2 + 1/4) \cosh^4(x/2)} -\frac{1}{\cosh^2(x/2)} \\
& s(r,x) = \frac{\tanh(x/2)}{2g^2(r,x)}\,,\quad & g(r,x) =\sqrt{r^2 + 1/4~ \tanh^2(x/2)},
\label{functions}\\
\label{functions2}
& \delta(x) = \delta_0 \cosh^4 (x/2)\,,\quad & \delta_0 = \frac{8 C}{c^2}\,.
\ea
Note that the scalar curvature is a negative constant globally for all solutions.
In the last equation $C$ and $c$ refer to the Casimir functions
\eqref{eq:scs7a} and \eqref{eq:scs7b} and accordingly the kink solution is
recovered for $\delta_0 = 0$ (recall the definition of $C$ in \eqref{eq:scs7a}
and the subsequent discussion). In the latter case
the eqs.~of motion derived from (\ref{4dgauged}), are satisfied,
\eqn{eom}{R_{\mu\nu}= 2 (F_{\mu \rho} F^\rho{}_\nu -\frac{1}{4} g_{\mu\nu}F^2 ) -3 g_{\mu\nu} }
with the gauge field strength
\begin{equation}
  \label{eq:fs3}
  F=\alpha e^0\wedge e^1+\beta e^0\wedge e^2+\alpha e^1\wedge e^3+\beta e^2\wedge e^3\,,
\end{equation}
where\footnote{As from now on we set $\delta_0=0$, the (useful)
relations $fg=(r^2+1/4)$ and $g=-fa\cosh^2{(x/2)}$ hold.}
\begin{equation}
  \label{eq:fs4}
  \alpha=\gamma r\tanh{(x/2)}\,,\quad\beta=-\gamma(r^2-\tanh^2{(x/2)}/4)\,,\quad\gamma=1/(2g^2\cosh{(x/2)})^2\,.
\end{equation}
To obtain \eqref{eq:fs3} the tetrad
\begin{equation}
  \label{eq:fs1}
  e^0=f(\extd t-a\extd z)\,,\quad e^1=\extd r/f\,,\quad e^2=g\extd x\,,\quad e^3=g\extd z/\cosh^2{(x/2)}\,
\end{equation}
has been employed and $F=\extd A$ has been used with
\begin{equation}
  \label{eq:fs0}
  \extd A=(\partial s/\partial r (r^2+1/4) +
  2rs)\extd r\wedge \extd t+\partial s/\partial x (r^2+1/4)\extd x\wedge \extd t\,.
\end{equation}
This implies an electric field $E=(\alpha,\beta,0)$ and a magnetic
field $B=(\beta,-\alpha,0)$ (the entries refer to
$1,2,3$-component, resp.). We can get $\ast F$ either by applying
Hodge to (\ref{eq:fs3}) or by replacing $\alpha\to\beta$ and
$\beta\to-\alpha$.
Note that $F\wedge\ast F=0$ but $F$ is not proportional to $\ast
F$, thus we have neither self duality nor anti-self duality.
The following physical consequences should be noted: both scalar
invariants built from $F$ vanish: $E^2-B^2=0$ and $E\cdot B=0$.
The energy-density turns out as
$H=\frac12(E^2+B^2)=\alpha^2+\beta^2=\gamma^2$ (for large $x$
energy behaves as $H\approx e^{-2x}/(r^2+1/4)^2$, for large $r$ it
reads $H \approx 1/(2r\cosh{(x/2)})^4$). The Poynting vector
$E\times B=-H(0,0,1)$ shows into negative $3$-direction. This is
the behavior expected from an ordinary electro-magnetic wave in
curved space. Note that the metric \eqref{metrica} is stationary
but not static, a property which also holds for the field
strength. One could call this field configuration a soliton
consisting of photons kept together by gravity: it is a regular
field configuration and energy is bounded for all values of $x,r$;
also, energy falls off for large $x,r$. Moreover, the nonvanishing
flux and its fall-off behavior explain why the Ricci tensor is
nontrivial but tends to the one of $AdS$ asymptotically
($x\to\infty$).

For very large values of the coordinate $x$, the metric looks like
\eqn{asympads4}{\extd s^2= (r^2+ \frac{1}{4})( -\extd t^2 + \extd
x^2 + e^{-x} \extd t \extd z ) + \frac{\extd r^2}{(r^2+
\frac{1}{4})} } whilst the gauge field (\ref{gaugefield}) is pure
gauge. For $x\to\infty$ this space is $AdS_4$. We will now discuss
the asymptotics in more detail in order to decide whether we have
$AAdS$ according to the definition given, for example, in
ref.~\cite{Skenderis:2002wp}.

\subsection{AdS asymptotics in D=2,3,4}\label{se:AAdS}

As there are several subtleties involved in the discussion of the
asymptotic behavior it is worthwhile to collect them at this
point. Because most of them arise already in the much simpler
framework of the 2D solutions we will study them first.

We recall that the Riemann tensor has $D^2(D^2-1)/12$ independent
components. Consequently, in D=2 the Riemann tensor may be
expressed solely in terms of the scalar curvature. Thus, it is
sufficient in D=2 to study the asymptotic behavior of the latter.
Now we encounter the first subtlety:

\paragraph{Definition of asymptotic region} The line element \eqref{ds4} suggests that the asymptotic region is located at $x\to\pm\infty$. One can then check that, according to the definition of $AAdS$ provided in ref.~\cite{Skenderis:2002wp}, the kink solution is $AAdS_2$. However, we have seen that $x\to\pm\infty$ does not correspond to the asymptotic region but rather to a Killing horizon. The asymptotic region of the analytically extended metric \eqref{eq:cs6} is reached for $X\to\pm\infty$. One can deduce immediately from \eqref{eq:cs7} that the metric is not $AAdS$. At least, not in the conformal frame employed there. This brings us to the second subtlety:

\paragraph{Definition of the conformal frame} Classical conformal invariance gives us the freedom to choose any non-singular conformal factor which we deem to be convenient. Actually, we may allow for singularities in the asymptotic region in order to get rid of the curvature singularity, but for finite $X$ the conformal factor must not vanish or diverge. Exploiting this freedom via eq.~\eqref{eq:cf} we have shown that it is possible to bring the metric to a form which asymptotically approaches $AdS$, cf.~eqs.~\eqref{eq:AdSkilling}-\eqref{eq:AdSgsr}. Thus, in a naive sense all our solutions asymptotically are $AdS$. However, there is another subtlety which appears to spoil the $AAdS$ property:

\paragraph{Next to leading order terms} It is mentioned in ref.~\cite{Skenderis:2002wp} that the Riemann tensor not only has to approach the one of $AdS$ space in the asymptotic region, but also that the next to leading order (NLO) terms should have a certain fall-off behavior and in particular an ordinary Taylor series in powers of the ``defining function'' should emerge. Let us study this in detail (for sake of definiteness we will restrict ourselves to $X\to\infty$). The asymptotic Killing norm in the frame implied by \eqref{eq:cf} reads $K=(\alpha/4)X^6+\mathcal{O}(X^4)$. With $\tilde{X}=(\alpha/3)X^3+\delta X$ (one over the ``defining function'' in the parlance of \cite{Skenderis:2002wp}) the asymptotic line element in conformal gauge ($\extd s^2_{\rm flat}=2\extd u\extd v$ with $\extd v=\extd u/2+\extd\tilde{X}/K(\tilde{X})$) reads (note that $\alpha>0$)
\begin{equation}
  \label{eq:AAdS}
  \extd s^2_{\rm asy} = \frac{9}{4\alpha} \tilde{X}^2\left(1+\mathcal{O}(\tilde{X}^{-2/3})\right) \extd s^2_{\rm flat}\,.
\end{equation}
As required, it has a second order pole in $1/\tilde{X}$ at the boundary. However, the NLO terms of \eqref{eq:AAdS} contain third roots and do not allow for an ordinary power series around $\tilde{X}=\pm\infty$. The only exception is the trivial case $C=c=0$ implying not only $AAdS_2$ but $AdS_2$.
Thus, although near the $AdS$ boundaries curvature is given by $r=-9/(2\alpha)+\mathcal{O}(\tilde{X}^{-2/3})$ spacetime is not $AAdS$ in the technical sense. If $\alpha=9/(4c)$ is chosen for matching continuously to the constant dilaton vacua (as explained below \eqref{eq:cf}) curvature at the boundary is given by $r=-2c$, in accordance with \eqref{rsol}-B.
It may well be that for some applications the appearance of third
roots in the NLO terms is not problematic, but it is emphasized again that in the strict sense none of our solutions is $AAdS$, despite of the correct asymptotic limit of the Riemann tensor and the line element.

\paragraph{Oxidation to D=3} 
In fact, most of the previous discussion ``oxidizes'' to
D=3. For a negative result it is sufficient to show that the scalar
curvature spoils the $AAdS$ property. A one-to-one repetition of
the line of reasoning given above, including all caveats
mentioned, leads to the conclusion that none of our solutions,
eq.~\eqref{eq:ct3Dmetric} below, is $AAdS_3$ in the strict sense because of
the NLO terms. What about being asymptotically $AdS$ in the weaker
sense explained above? 
We present a general argument why the metric can be brought
into a form which asymptotically approaches $AdS_3$: because the
Cotton tensor \eqref{cottontensor} vanishes globally the line
element \eqref{eq:3Dmetric} is conformally flat. But since $AdS_3$
is conformally flat as well it is possible by two consecutive
conformal transformations to bring the line element to $AdS_3$.
However, the conformal factor will be singular at Killing horizons
of \eqref{eq:3Dmetric}. Thus, by excluding singularities in the
conformal factor everywhere except in the asymptotic region it is
only possible to transform \eqref{eq:3Dmetric} to a form which
asymptotically, i.e., for $X\to\pm\infty$, approaches $AdS_3$. Therefore it is tempting to use the line element
\begin{equation}
  \label{eq:ct3Dmetric}
  \extd s^2 = e^Q \left(2C\extd u^2+2\extd u \extd X - (X^2-c)\extd u\extd r -\extd r^2\right)\,,
\end{equation}
with $Q$ as defined in \eqref{eq:cf}. At first glance prospects look promising:
in D=3 the Riemann tensor is fully determined by specifying the
trace free Ricci tensor and the curvature scalar, and the latter
tends to a negative constant asymptotically. Additionally, in 3D
the Cotton tensor $C^{\mu\nu}$ has as many algebraically
independent components as the trace free Ricci tensor, namely 5
\cite{Garcia:2003bw}, and the eqs.~of motion imply $C^{\mu\nu}=0$.
However, $C^{\mu\nu}$ contains one additional derivative as
compared to the trace free Ricci tensor and thus less information.
Therefore, it is conceivable that some integration constants are
present in the trace free Ricci tensor while $C^{\mu\nu}=0$.
Indeed, this turns out to be the case for the line element
\eqref{eq:ct3Dmetric}. Consequently, our 3D solutions \eqref{eq:ct3Dmetric} are not even
asymptotically $AdS$ in a weak sense. It seems likely that the conformal factor which brings \eqref{eq:ct3Dmetric} to a form which is asymptotically $AdS_3$ has to depend not only on the dilaton $X$, but also on the other two coordinates.

\paragraph{Consequences for D=4}We have mentioned in the previous
section that the 4D analogue of the BPS solution asymptotically
approaches $AdS_4$, so it could well be that it is $AAdS_4$.
However, again most of the caveats mentioned above apply: In
particular, the ``asymptotic region'' $x\to\pm\infty$ has been
employed, but since neither in D=2 nor in D=3 this corresponds to
the true asymptotic region it remains an open question whether
this is the case in D=4. Moreover, while the limit $x\to\pm\infty$
of the Riemann tensor undoubtedly is consistent with $AdS_4$ the
NLO terms have not been checked. Finally, because the scalar
curvature is constant for all solutions \eqref{metrica} and
because the Weyl tensor $C^\mu{}_{\nu\sigma\tau}$
vanishes\footnote{It is emphasized that this is not true for the
index position $C_{\mu\nu\sigma\tau}$. However, the index position
used in the main text appears to be preferable because only then
the Weyl tensor is conformally invariant.} in the limit
$r\to\infty$ also the non-BPS solutions exhibit nice asymptotic
features: they approach $AdS_4$ in that limit and the gauge field
becomes pure gauge. It could be worthwhile to extend our analysis
of the asymptotic behavior in D=4, in particular, to resolve what
is actually meant by ``asymptotic''. This will be left for future
work.

\subsection{Lifting to M theory}

When lifting this 4D solution to M theory, we use the prescription
given in \cite{Cvetic:1999xp}. We define the quantities (note that
$\Sigma_i \mu_i^2=1$), \eqn{mus}{\mu_1=\sin\theta\,,\; \mu_2 =
\cos\theta \sin\varphi\,, \; \mu_3= \cos\theta
\cos\varphi\sin\psi\,,\;\; \mu_4= \cos\theta \cos\varphi
\cos\psi\,.} So that the seven sphere can be written as
\eqn{s7}{\extd\Omega_7^2= \sum_i [(\extd\mu_i)^2 +\mu_i^2
\extd\phi_i^2],\;\; i=1,..,4} The 11D metric and four form read
(the star denotes the 4D Hodge star), \eqn{eleven}{\extd s_{11}^2
= \extd s_4^2 +  4 \sum_i [(\extd\mu_i)^2 +
  \mu_i^2 (\extd\phi_i -\frac{1}{2} \Afour_t \extd t)^2] }
\eqn{fourform}{F_4= -3 \epsilon_4  -\sum_i [\extd(\mu_i^2)\wedge (\extd\phi_i
  + \Afour_t \extd t)]\wedge {}^*F_2.}
So, we see that for large values of the coordinate $x$ the field strength
$F_2$ in \eqref{eq:fs3} vanishes and the
configuration looks like M2 branes. This asymptotic metric, of the form $AdS_4 \times S^7$ is
being deformed by the presence of the fibration between the four
dimensional part $\extd s^2_4$ and the internal space, a deformed $S^7$. To compensate for this change
and satisfy the Einstein eqs., the four form field strength gets the second factor in
eq.~(\ref{fourform}).

The fact that the metric may be $AAdS$ opens the possibility of
studying $AdS/CFT$ aspects of the 11D solution. So, this solution
is representing a dual to a 3D conformal theory, that is being
deformed by the insertion of an operator or a VEV in the UV, this
insertion or VEV also breaks part of the SUSY. We could compute
correlation functions by following the prescription in
\cite{Skenderis:2002wp}. In order to explore the above mentioned
issues, one should first write the metric in the form indicated in
ref.~\cite{Skenderis:2002wp}. It appears to be difficult to find
the explicit transformation, though it should be possible in
principle.\footnote{Mauro Brigante has found a change of variables
that leaves the metric (\ref{asympads4}) in the form $g =
\phi(t,x,r,z) \eta$ where $\eta$ is the flat metric.} We leave the
$AdS/CFT$ analysis for a future work.

To end this section it would be remark to note the following
points. On the one hand it should be interesting to make a full
mapping between the 4D gauged SUGRA in eq.~(\ref{4dgauged}) and
the 2D SUGRA in (\ref{eq:scs13main}). If this mapping is possible
for any solution of (\ref{eq:scs13main}), then, the matter fluxes
that stabilize the kink as discussed in section \ref{se:2} could
be identified with the gauge fields in the 4D context and by
lifting, with the M theory four form field strength. On the other
hand, the mass formula discussed in section \ref{se:2} should be
related to the ADM mass of the 4D or 11D solution. There should be
a relation between the Hawking temperature discussed in section
\ref{se:ht} from the 2D viewpoint and some observable in 4D gauged
SUGRA or M theory, which will be addressed in a future work as well.

\section{Discussion and Conclusions}

One of our main results was the construction of a supersymmetric
extension of the dimensionally reduced gravitational Chern-Simons
term
\begin{equation}
  \label{ceq:scs13main}
\Stext{SUCS} = -\frac{1}{8\pi^2} \intd{\diff{^2 x}\sqrt{-g}}
\Bigl( \tilde{r} F + F^3 + \Sigma^2 - F^2 \Delta \Bigr)
\end{equation}
with SUSY transformations
\begin{align}
  \delta {e_m}^a &= - 2i (\ve\gamma^a \psi_m)\ , \label{ceq:scs12.3main}\medsp
  \label{ceq:scs12.6main}
  \delta \A_m &= - 2\ve \gamma_* \psi_m\ ,\medsp
  \label{ceq:scs12.4main}
  \delta \psi_{m \alpha} &= - \hat{D}_m \ve_\alpha \;.
\end{align}
We have also shown that this model is not equivalent to the reduction of the 3D SUCS, as the fermionic transformations and potential differ. We will address this model in future work.

We have recalled how to obtain all classical solutions locally and
globally, namely by casting \eqref{ceq:scs13main} into first order
form and exploiting some of the features of graded
Poisson Sigma Models. By the use of gPSM techniques global
properties of the kink solution including its BPS characteristics have
been discussed rigorously. These results were compared with the higher
dimensional perspective.

The thermodynamical behavior of SUCS solutions is quite
non-trivial. Here we summarize briefly the various phases,
supposing $c>0$ (in brackets the global solution in the notation
of ref.~\cite{Grumiller:2003ad} is provided):\footnote{The case
$c\leq 0$ can be studied as well, but it is less interesting. The
only geometries arising are {\bf B0}, {\bf B1a}, {\bf B1b} and
{\bf B2a}, i.e., at most two horizons are possible.}
\begin{itemize}
\item $M<0$: no horizon ({\bf B0})
\item $M=0$: two extremal horizons, BPS solution, Hawking temperature vanishes, entropy\footnote{Entropy may be calculated by various methods \cite{Gegenberg:1995pv} and in our convention reads $S=2\pi|X_h|$, where $X_h$ is the value of the dilaton field at a horizon given by \eqref{eq:killing0}.} is proportional to $\sqrt{c}$ ({\bf B2b})
\item $0<\sqrt{2M}<c/3$: four horizons, inner ones: positive specific heat ({\bf B4})
\item $\sqrt{2M}=c/3$: four horizons, Hawking-Page like transition on inner horizons: specific heat diverges ({\bf B4})
\item $c/3<\sqrt{2M}<c/2$: four horizons, inner horizons: negative specific heat ({\bf B4})
\item $\sqrt{2M}=c/2$: two non-extremal and one extremal horizon, Hawking temperature and entropy vanish on inner (extremal) horizon ({\bf B3})
\item $\sqrt{2M}>c/2$: two horizons ({\bf B2a})
\end{itemize}

The specific heat on outer horizons is always positive. It is
emphasized that the simple thermodynamical relations presented in
this paper are but the first step to a thorough analysis.  In
particular, the presence of several horizons with different
surface gravities has to be taken into account properly. Also, a
better (microscopical) understanding of the entropy -- e.g.~for
the BPS solutions where it grows like the square-root of the
$U(1)$ charge -- would be desirable. Finally, the Hawking-Page
like transition implied by the pole in \eqref{eq:specificheat}
should be studied in more detail. Especially the last point could
be rewarding to pursue due to its relevance for $AdS$/CFT
(cf.~sect.~3.2 in ref.~\cite{Witten:1998qj} and sect.~2.3 in
ref.~\cite{Witten:1998zw}).

In this context we have addressed the asymptotic behavior of the
metric and the Riemann tensor in D=2,3 and 4 thereby revealing
promising features: in a convenient conformal frame all solutions
in D=2 and in D=3 asymptotically tend to $AdS$. However, the next to
leading order terms contain third roots which seem to spoil $AAdS$
in the strict sense \cite{Skenderis:2002wp}. Further subtleties
have been discussed and their resolution in D=4 remains as an open
task for future work. In particular, it has to be clarified what
is actually the asymptotic region in D=4 and in D=11 (obtained by
lifting the 4D solutions to M theory).

Let us finally address generalizations of our theory that include coupling
to matter. In general, the theory ceases to be a topological one
and allows for physical scattering processes. Nevertheless, an
exact path integral quantization of geometry, auxiliary fields and
ghosts may be performed, thus providing a generating functional
for Green functions depending solely on the matter fields and on
external sources \cite{Bergamin:2004us}. It could be of interest
to apply the general results in that ref.~to the present case and
to study scattering in various phases of the model, starting for
simplicity with $M=0$. However, dimensional reduction makes it
natural to assume a non-trivial coupling of matter to the dilaton
field. Though it has been shown recently that the basic properties
of the quantization procedure in \cite{Bergamin:2004us} are
retained together with non-minimal coupling
\cite{Bergamin:2004aw}, this case has not yet been worked out in
detail. Moreover, non-trivial couplings of the matter fields to
the new scalar field $Y$ could yield an even richer structure
which will be worthwhile to study.

\acknowledgments

\noindent A.I. acknowledges the kind hospitality of the Institute
for Theoretical Physics of Leipzig while part of this paper was
conceived. D.G. acknowledges the kind hospitality of the Center
for Theoretical Physics at MIT where some of the bosonic topics
have been addressed. We thank Roman Jackiw for calling our
attention to this problem and Mauro Brigante, Stanley Deser,
Daniel Freedman, Umut Gursoy, Amihay Hanany, Matt Headrick,
Wolfgang Kummer, Hong Liu,  Rishi Sharma, Dima Vassilevich and
Barton Zwiebach for valuable and enjoyable discussions.

\begin{appendix}

\section{Notations and conventions}
\label{app:A}
The conventions used in the first order formulation in section \ref{se:1} are identical to
\cite{Ertl:2000si}, where additional explanations can be found.

Indices chosen from the Latin alphabet are commuting (lower case)
or generic (upper case), Greek indices are anti-commuting.
Holonomic coordinates are labeled by $M$, $N$, $O$ etc.,
anholonomic ones by $A$, $B$, $C$ etc., whereas $I$, $J$, $K$
etc.\ are general indices of the gPSM. The index $\phi$ is used to
indicate the dilaton component of the gPSM fields (note that in
the main text we have used $X$ instead for sake of compatibility
with the bosonic literature, while here we use $\phi$ for sake of
compatibility with the SUSY literature):
  \begin{align}
    X^\phi &= \phi & A_\phi &= \omega \\
    X^Y &= Y & A_Y &= \Aapp
  \end{align}

The summation convention is always $NW \rightarrow SE$, e.g.\ for a
fermion $\chi$: $\chi^2 = \chi^\alpha \chi_\alpha$. Our conventions are
arranged in such a way that almost every bosonic expression is transformed
trivially to the graded case when using this summation convention and
replacing commuting indices by general ones. This is possible together with
exterior derivatives acting \emph{from the right}, only. Thus the graded
Leibniz rule is given by
\begin{equation}
  \label{eq:leibniz}
  \mbox{d}\left( AB\right) =A\mbox{d}B+\left( -1\right) ^{B}(\mbox{d}A) B\ .
\end{equation}

In terms of anholonomic indices the metric and the symplectic $2 \times 2$
tensor are defined as
\begin{align}
  \eta_{ab} &= \left( \begin{array}{cc} 1 & 0 \\ 0 & -1
  \end{array} \right)\ , &
  \epsilon_{ab} &= - \epsilon^{ab} = \left( \begin{array}{cc} 0 & 1 \\ -1 & 0
  \end{array} \right)\ , & \epsilon_{\alpha \beta} &= \epsilon^{\alpha \beta} = \left( \begin{array}{cc} 0 & 1 \\ -1 & 0
  \end{array} \right)\ .
\end{align}
The metric in terms of holonomic indices is obtained by $g_{mn} = e_n^b e_m^a
\eta_{ab}$ and for the determinant the standard expression $e = \det e_m^a =
\sqrt{- \det g_{mn}}$ is used. The volume form reads $\epsilon = \half{1}
\epsilon^{ab} e_b \wedge e_a$; by definition $\ast \epsilon = 1$.

The $\gamma$-matrices are used in a chiral representation:
\begin{align}
\label{eq:gammadef}
  {{\gamma^0}_\alpha}^\beta &= \left( \begin{array}{cc} 0 & 1 \\ 1 & 0
  \end{array} \right) & {{\gamma^1}_\alpha}^\beta &= \left( \begin{array}{cc} 0 & 1 \\ -1 & 0
  \end{array} \right) & {{\gthree}_\alpha}^\beta &= {(\gamma^1
    \gamma^0)_\alpha}^\beta = \left( \begin{array}{cc} 1 & 0 \\ 0 & -1
  \end{array} \right)
\end{align}

Covariant derivatives of anholonomic indices with respect to the
geometric variables $e_a = \extd x^m e_{am}$ and $\psi_\alpha =
\extd x^m \psi_{\alpha m}$ include the 2D spin-connection one form
$\omega^{ab} = \omega \epsilon^{ab}$. When acting on lower indices
the explicit expressions read ($\half{1} \gthree$ is the generator
of Lorentz transformations in spinor space):
\begin{align}
\label{eq:A8}
  (D e)_a &= \extd e_a + \omega {\epsilon_a}^b e_b & (D \psi)_\alpha &= \extd
  \psi_\alpha - \half{1} {{\omega \gthree}_\alpha}^\beta \psi_\beta
\end{align}

Light-cone components are very convenient. As we work with spinors in a
chiral representation we can use
\begin{align}
\label{eq:Achi}
  \chi^\alpha &= ( \chi^+, \chi^-)\ , & \chi_\alpha &= \begin{pmatrix} \chi_+ \\
  \chi_- \end{pmatrix}\ .
\end{align}
For Majorana spinors upper and lower chiral components are related
by $\chi^+ = \chi_-$, $ \chi^- = - \chi_+$, $\chi^2 = \chi^\alpha
\chi_\alpha = 2 \chi_- \chi_+$. Vectors in light-cone coordinates
are given by
\begin{align}
\label{eq:A10}
  v^{++} &= \frac{i}{\sqrt{2}} (v^0 + v^1)\ , & v^{--} &= \frac{-i}{\sqrt{2}}
  (v^0 - v^1)\ .
\end{align}
The additional factor $i$ in \eqref{eq:A10} permits a direct identification of the light-cone components with
the components of the spin-tensor $v^{\alpha \beta} = \frac{i}{\sqrt{2}} v^c \gamma_c^{\alpha
  \beta}$. This implies that $\eta_{++|--} = \eta_{--|++} = 1$
and $\epsilon_{--|++} = - \epsilon_{++|--} = 1$. The
$\gamma$-matrices in light-cone coordinates become
\begin{align}
\label{eq:gammalc}
  {(\gamma^{++})_\alpha}^\beta &= \sqrt{2} i \left( \begin{array}{cc} 0 & 1 \\ 0 & 0
  \end{array} \right)\ , & {(\gamma^{--})_\alpha}^\beta &= - \sqrt{2} i \left( \begin{array}{cc} 0 & 0 \\ 1 & 0
  \end{array} \right)\ .
\end{align}

\section{First order formulation of SUCS}\label{app:B}

For details on notations Appendix~\ref{app:A} may be consulted.
Note that the dilaton is denoted by $\phi$ in the appendices and
not by $X$ for sake of backward compatibility with the literature
on dilaton SUGRA. With the prepotential \eqref{eq:scs1} the explicit expressions for the Poisson tensor
are \cite{Bergamin:2003am}:
\begin{align}
  \label{eq:scs2a}
  P^{a\phi} &= X^b\epsilon_b{}^a\medsp
  \label{eq:scs2b}
  P^{\alpha\phi} &= -\frac12 \chi^\beta\gthree_{\beta}{}^\alpha\medsp
  \label{eq:scs2}
  P^{ab} &= \epsilon^{ab} \Bigl(\half{1} (\phi \B - \phi^3) + \inv{8} \chi^2
  \Bigr)\medsp
  \label{eq:scs3}
  P^{\alpha b} &= - \frac{i}{2} \phi (\chi \gamma^b)^\alpha\medsp
  \label{eq:scs4}
  P^{\alpha \beta} &= - 2i X^c \gamma_c^{\alpha \beta} + (\phi^2 - \B)
  \gthree^{\alpha \beta}\medsp
  \label{eq:scs5}
  P^{\B I} &\equiv 0
\end{align}
It should be noted that it does not have full rank. The dimension
of the kernel is two and thus two Casimir functions\footnote{The
name ``Casimir'' is justified because, introducing the
Schouten-Nijenhuis bracket $\{X^I,X^J\}=P^{IJ}$, a Casimir function
fulfills $\{X^I,C\}=P^{IJ}\partial C/\partial X^J=0$ for all $I$.
On a sidenote, the non-linear Jacobi identity \eqref{eq:nijenhuis}
is then seen to be a simple consequence of the ``ordinary'' Jacobi
identity for this bracket.} exist, related to conserved mass and
charge.

The first-order action (\ref{eq:gPSM}) becomes
\begin{multline}
\label{eq:scs6}
  \Stext{FO} = \int_{\mathcal{M}_2} \Bigl( \phi \extd \omega + \B \extd \Aapp + X^a De_a + \chi^\alpha D
  \psi_\alpha + \epsilon \bigl(\half{1} (\phi \B - \phi^3) + \inv{8} \chi^2
  \bigr) \medsp
  - \frac{i}{2} \phi (\chi \gamma^a e_a \psi) + i X^a (\psi \gamma_a \psi) -
  \half{1}(\phi^2 - \B) (\psi \gthree \psi) \Bigr)\ .
\end{multline}
The eqs.~of motion
\begin{align}
\label{eq:gPSMeom1}
  \extd X^I + P^{IJ} A_J &= 0\ ,\medsp
\label{eq:gPSMeom2}
  \extd A_I + \half{1} (\partial_I P^{JK}) A_K \wedge A_J &= 0\ ,
\end{align}
can be found explicitly in the main text, \eqref{eq:gPSMeom3.1}-\eqref{eq:neweom2}. The change of notation between appendix and main text should be noted, e.g.~the dilaton field $\phi$ is denoted by $X$ in the main text; other changes are summarized in footnote \ref{fn:ref}.

Elimination of $X^a$ and bosonic torsion yields (eq.\ (2.38) in
\cite{Bergamin:2003am})
\begin{multline}
  \label{eq:scs7}
  \Stext{SO} = \intd{\diff{^2 x}} e \Bigl(\half{1} \tilde{R} \phi + (\chi
  \tilde{\sigma}) - \epsilon^{mn} (\B \partial_n \Aapp_m) + \half{1} (\phi \B -
  \phi^3) \medsp + \inv{8} \chi^2  + \half{i} \phi \epsilon^{mn} (\chi \gamma_n
  \psi_m) + \half{1} (\phi^2 - \B) \epsilon^{mn} (\psi_n \gthree \psi_m) \Bigr)
\end{multline}
with the SUSY covariant scalar curvature and its supersymmetric partner
\begin{align}
  \label{eq:scs8}
  \tilde{R}&=2 \ast \extd \tilde{\omega } = 2 \epsilon^{mn} \partial_n
  \tilde{\omega}_m\ , \medsp
  \label{eq:scs9}
  \tilde{\sigma}_\alpha &= \ast (\tilde{D} \psi)_\alpha =
  \epsilon^{mn} \bigl( \partial_n \psi_{m\alpha} + \half{1} \tilde{\omega}_n (\gthree
  \psi_m)_\alpha \bigr)\ ,\medsp
  \label{eq:scs10}
  \tilde{\omega}_a &= \epsilon^{mn} \partial_n e_{ma} - i \epsilon^{mn} (\psi_n
  \gamma_a \psi_m)\ .
\end{align}
As in the bosonic theory $\B$ in \eqref{eq:scs7} is a Lagrange multiplier,
eliminating $\phi$ according to
\begin{equation}
  \label{eq:scs11}
  \phi = \epsilon^{mn} (2\partial_n \Aapp_m + \psi_n \gthree \psi_m) =: \tilde{F}\ .
\end{equation}
Then the action \eqref{eq:scs7} reduces to
\begin{equation}
  \label{eq:scs12}
  \Stext{SO} = \half{1}\intd{\diff{^2 x}} e\Bigl(\tilde{R} \tilde{F} -
  \tilde{F}^3 + 2(\chi
  \tilde{\sigma}) + \inv{4} \chi^2 \medsp + i \tilde{F} \epsilon^{mn} (\chi \gamma_n
  \psi_m) + \tilde{F}^2 \epsilon^{mn} (\psi_n \gthree \psi_m) \Bigr)\ .
\end{equation}
From the gPSM symmetries and the elimination conditions for $X^a$
and torsion one obtains as SUSY transformations in \eqref{eq:scs7}
(cf.\ eqs.\ (2.40)-(2.43) in \cite{Bergamin:2003am})
\begin{align}
  \delta \phi &= \half{1} (\chi \gthree \ve)\ , \label{eq:scs12.1}\medsp
  \delta \chi^\alpha &= - 2i \epsilon^{mn} \bigl( \partial_n \phi + \half{1}
  (\chi \gthree \psi_n) \bigr)  (\ve \gamma_m)^\alpha - \bigl(\phi^2 -\B \bigr) (\ve \gthree)^\alpha\ , \label{eq:scs12.2}\medsp
  \label{eq:scs12.5}
  \delta \B &\equiv 0\,,\medsp
  \delta {e_m}^a &= - 2i (\ve \gamma^a \psi_m)\ , \label{eq:scs12.3}\medsp
  \delta \psi_{m \alpha} &= - (\tilde{D} \ve)_{m\alpha} - \frac{i}{2} \phi(\gamma_m
  \ve)_\alpha \,,
\label{eq:scs12.4}\medsp \label{eq:scs12.6}
 \delta \Aapp &= - \ve \gthree \psi\,.
\end{align}
It is emphasized that the choice of the prepotential
\eqref{eq:scs1} not only determines the bosonic potential
\eqref{eq:scs1a} but also the transformation law of $\Aapp$ in
\eqref{eq:scs12.6}. Because $\Aapp$ stems from a component of the
Dreibein this is really the correct transformation law. Also all
other transformations are as expected.

Elimination of $\chi$ from \eqref{eq:scs12} yields
\begin{equation}
  \label{eq:chiel}
\chi_\alpha = -4 \tilde{\sigma}_\alpha - 2i \tilde{F} \epsilon^{mn}
(\gamma_n \psi_m)_\alpha = -4\epsilon^{mn}(\hat{D}_n\psi_m)_\alpha\,,
\end{equation}
where $\hat{D}$ is defined as in \eqref{hatD}. This allows to eliminate $\tilde{\sigma}_\alpha$ in terms of $\chi_\alpha$ and a term proportional to $\tilde{F}$. Insertion into the action (\ref{eq:scs12}) and changing to the notation\footnote{\label{fn:ref} Note especially $\tilde{R}=-r$,
$\tilde{F}=f+\epsilon^{mn}\psi_n\gthree\psi_m$, $\Aapp=\A/2$ and $e=\sqrt{-g}$.}
of \cite{Guralnik:2003we} after multiplication with
$1/(4\pi^2)$ establishes
\begin{equation}
  \label{eq:scs13}
\Stext{SO} = -\frac{1}{8\pi^2} \intd{\diff{^2 x}}\sqrt{-g} \Bigl(
r\tilde{F} + \tilde{F}^3 + \frac14 \chi^2 - \tilde{F}^2\epsilon^{mn}\psi_n\gthree\psi_m \Bigr)\,.
\end{equation}
The transformations \eqref{eq:scs12.3}-\eqref{eq:scs12.6} are
unchanged except for \eqref{eq:scs12.4} which now reads instead
\begin{equation}
  \label{eq:scs12.4new}
\delta \psi_{m \alpha} = - (\tilde{D} \ve)_{m \alpha} - \half{i}
\tilde{F} (\gamma_m \ve)_\alpha = -(\hat{D}_m\epsilon)_\alpha\ .
\end{equation}
The action presented in the main text, Eq.~\eqref{eq:scs13main}, is a direct consequence of \eqref{eq:scs13} together with \eqref{eq:chiel}.

The action \eqref{eq:scs13} can be obtained from a superspace formulation as
well. This can be seen from a result of ref.\ \cite{Bergamin:2003am} where it was shown
that $N=(1,1)$ gPSM models with vanishing bosonic torsion can be mapped onto
superspace actions of the model of Howe \cite{Howe:1979ia}. In the former case the
independent spin connection, $X^a$ and $\chi^\alpha$ can be eliminated yielding a
theory formulated in terms of dilaton, zweibein and gravitino and completely
determined by the prepotential $u(\phi)$. The superspace fomulation consists
of one multiplet $(e_m^a, \psi_m^\alpha, \mathcal{A})$, where $\mathcal{A}$ is the auxiliary field.
The action is determined by a function $\mathcal{F}(S)$ with $S$ being a scalar
superfield that has $\mathcal{A}$ as its lowest component (cf.\ \cite{Howe:1979ia}, we adopt the
notation of \cite{Bergamin:2003am}). The two theories are equivalent with the identification
\begin{align}
\mathcal{A} &= - \half{u'(\phi)}\ , & \mathcal{F}(\mathcal{A}) = \half{1}
\Bigl( u\bigl(\phi(\mathcal{A})\bigr) - \phi(\mathcal{A})
u'\bigl(\phi(\mathcal{A})\bigr)\Bigr)\ .
\end{align}
Although our model is not equivalent to the one of ref.\ \cite{Bergamin:2003am} due to the
additional fields $Y$ and $A$, the identification still turns out to be
straightforward. The condition $\mathcal{A} = - \phi$ obviously is globally defined and
independent of $Y$. This does not apply to $\mathcal{F}(\mathcal{A}) = -1/2 (\mathcal{A}^2 + Y)$, but it is
easily seen e.g.\ from \eqref{eq:scs7} that the $Y$-dependent terms simply reproduce
the constraint \eqref{eq:scs11}. Thus the action \eqref{eq:scs13} is---up to overall
factors---equivalent to the superspace action \eqref{eq:howe1}
if the auxiliary field $\mathcal{A}$ is interpreted as the dual field strength
according to \eqref{eq:scs11}.

\section{Conformal transformations}\label{app:C}
Dilaton dependent (super-)conformal transformations \cite{Howe:1979ia,Park:1993sd}
\begin{equation}
  \label{eq:ctd}
  g\to e^{Q(X)}g
\end{equation}
on the world sheet correspond---up to a redefinition of the gravitino---to
target space diffeomorphisms at the level of the gPSM \cite{Bergamin:2003am}. They introduce (or change) the second potential $Z$ in the bosonic part of the action (cf.~the Eq.~in footnote \ref{fn:st}), which is related to the conformal factor by means of
\begin{equation}
  \label{eq:Z}
  Z = \frac{\extd}{\extd X} Q\,.
\end{equation}
With the redefinition
\begin{equation}
  \label{eq:ctredef}
  \extd\tilde{X}=e^Q\extd X
\end{equation}
the transformed line element attains the form
\begin{equation}
  \label{eq:ctle}
  \extd\tilde{s}^2=2\extd u\extd \tilde{X} + \underbrace{e^Q K(X;C,c)}_{:=\tilde{K}(\tilde{X};C,c)}\extd u^2\,,
\end{equation}
where $X$ in the Killing norm may be expressed as a function of $\tilde{X}$ by integrating \eqref{eq:ctredef}. It should be noted that the derivative of the Killing norm
\begin{equation}
  \label{eq:ctsg}
  \frac{\extd}{\extd\tilde{X}}\tilde{K}=e^{-Q}\frac{\extd}{\extd X} \left(e^Q K \right) = K Q^\prime + \frac{\extd}{\extd X} K
\end{equation}
is conformally invariant only at Killing horizons $\tilde{K}=0=K$. Consequently, Hawking temperature as derived naively from surface gravity is conformally invariant.

For simplicity the focus will be on conformal factors monomial in the dilaton,
\begin{equation}
  \label{eq:ctmono}
  e^Q=\alpha X^\beta\quad \rightarrow \quad Z=\frac{\beta}{X}\,,
\end{equation}
with $\beta\neq -1$.
Strictly speaking, one should require positivity, but as the explicit
examples below will exclusively be restricted to even $\beta$ and
positive $\alpha$ we can avoid the introduction of absolute values in
\eqref{eq:ctmono}. The transformed Killing norm \eqref{eq:killing},
\begin{equation}
  \label{eq:ctkilling}
  \tilde{K} = \alpha \left(\frac{\beta+1}{\alpha}\tilde{X}\right)^{\beta/(\beta+1)}\left[2C+\frac14 \left(\left(\frac{\beta+1}{\alpha}\tilde{X}\right)^{2/(\beta+1)}-c\right)^2\right]\,,
\end{equation}
leads to the transformed scalar curvature
\begin{multline}
  \label{eq:ctr}
  \tilde{r}=\alpha\Big[2\big(C+\frac{c^2}{8}\big)\frac{\beta}{(\beta+1)^2}\big(\frac{\beta+1}{\alpha}\big)^{\beta/(\beta+1)} \tilde{X}^{-(\beta+2)/(\beta+1)} \\
+\frac{c}{2}\big(\frac{\beta+1}{\alpha}\big)^{(\beta+2)/(\beta+1)}\frac{\beta+2}{(\beta+1)^2}\tilde{X}^{-\beta/(\beta+1)}\\
- \frac34 \big(\frac{\beta+1}{\alpha}\big)^{(\beta+4)/(\beta+1)}\frac{\beta+4}{(\beta+1)^2} \tilde{X}^{-(\beta-2)/(\beta+1)} \Bigr]\,,
\end{multline}
which is slightly more complicated than \eqref{eq:cs7}.
The term in the last line provides curvature of the ground state geometry $C=c=0$. There are now a couple of interesting special cases.

\paragraph{Minkowski ground state} For $\beta=-4$ curvature of the ground state vanishes identically. This can happen only if the ground state geometry is either Minkowski or Rindler space. To achieve the latter the ground state Killing norm has to be linear in $\tilde{X}$, which is not possible. Rather, the Killing norm is constant and thus one obtains a Minkowskian ground state. Incidentally, it should be noted that in the original frame the ground state geometry is given by {\bf B1b} (see fig.~\ref{fig:CPs}). Thus, the singularity of the conformal factor at $X=0$ is responsible for the absence of horizons in the ground state solution in the Minkowski ground state frame. Minkowski ground state frames are very natural for energy definitions as the notion of ``ADM mass'' makes sense. As can be deduced from \eqref{eq:ctkilling} the ADM mass, up to a scale factor, will be given by $C+c^2/8$. Thus -- exactly like for the extremal RN BH -- all BPS kinks have the same BPS mass $C=0$ but different ADM masses, depending on the charge (cf.~footnote \ref{fn:BPS}). So is there anything to be said against the use of this frame? The main objection is not the fact that the conformal factor is singular at $X=0$, because this could be worked around by allowing binomial (or more complicated) conformal factors which only asymptotically behave monomial, but the fact that the asymptotic region is geodesically incomplete. While the application of this frame is excellent for purposes which involve a ``far field approximation'' like for Schwarzschild or the RN BH, it becomes less attractive if the asymptotic region turns out to be not asymptotic after all. Thus, unfortunately, for the model under consideration this frame is not very useful.

\paragraph{Dual frame} For $\beta=-2$ the term in the second line of \eqref{eq:ctr} vanishes. The name ``dual frame'' has been chosen because $\tilde{X}\propto 1/X$ (we recall that on-shell $X=f$). Thus, for ``strong coupling'' (large values of the dual field strength $f$ and thus of the dilaton field $X$) in the original frame we encounter small values of the transformed dilaton field $\tilde{X}$.

\paragraph{Original frame} Trivially, for $\beta=0$ one remains in the original frame. We would like to pinpoint that only in this frame the contribution in the first line of \eqref{eq:ctr} vanishes. This is the ``somewhat counter intuitive feature'' discussed below Eq.~\eqref{eq:cs7} and thus we see that it is truly an artifact of that frame.

\paragraph{AdS ground state} For $\beta=2$ ground state geometry has constant curvature $r_0$. Because $\alpha$ has to be positive $r_0<0$ and thus the ground state is $AdS$. This is a very nice behavior because also the SUSY preserving constant dilaton vacua are $AdS$ and thus in such a frame the kink solution -- if it existed -- could be matched in the asymptotic region with the constant dilaton vacua {\em without induced matter fluxes}. Actually, this is very much in the spirit of the original work \cite{Guralnik:2003we} where no matter fluxes are present because there geometry does not extend over the horizons. As this case is the most relevant for our purposes an explicit expression for the Killing norm will be provided,
\begin{equation}
  \label{eq:AdSkilling}
  \tilde{K} = \alpha \left(\frac{3}{\alpha}\tilde{X}\right)^{2/3}\left[2C+\frac14 \left(\left(\frac{3}{\alpha}\tilde{X}\right)^{2/3}-c\right)^2\right]\,,
\end{equation}
and for the associated scalar curvature
\begin{equation}
  \label{eq:AdSgs}
  \tilde{r}=\alpha\Big[\big(C+\frac{c^2}{8}\big)\frac{4}{9}\big(\frac{3}{\alpha}\big)^{2/3} \tilde{X}^{-4/3}
+\frac{2c}{9}\big(\frac{3}{\alpha}\big)^{4/3}\tilde{X}^{-2/3}\Bigr] + r_0\,,
\end{equation}
containing the ground state curvature
\begin{equation}
  \label{eq:AdSgsr}
  r_0 = - \frac{9}{2\alpha} \,.
\end{equation}
Note that the original asymptotic region $X=\pm\infty$ still corresponds to the transformed asymptotic region $\tilde{X}=\pm\infty$. Thus, any conformal factor which asymptotically behaves monomially as in \eqref{eq:ctmono} with $\beta=2$ will exhibit the desirable feature of a ground state geometry which asymptotically is $AdS$.

\paragraph{Final remark} As can be argued on general grounds a monomial conformal factor \eqref{eq:ctmono} will always destroy the kink solution (unless $\beta=0$). This can be checked more explicitly by noticing that for $\beta<-1$ the origin is mapped to infinity and thus the kink may not pass through it. On the other hand, for $\beta>-1$ the exponent of $\tilde{X}$ in the first term in \eqref{eq:ctr} is always negative. Thus, a curvature singularity is introduced at the origin (unless $\beta=0$ where this term vanishes identically) and again the kink solution may not pass through. Therefore, in the main text a suitable conformal factor binomial in $X$ is studied which still retains the favorable features of the last case above.

\end{appendix}



\providecommand{\href}[2]{#2}\begingroup\raggedright\endgroup

\end{document}